\documentclass[showkeys,showpacs,amsmath,amssymb,a4paper,12pt]{revtex4}
\usepackage{amsmath}
\usepackage{amssymb}
\usepackage{amsmath,amsthm,amssymb,amscd}
\usepackage{latexsym}
\usepackage{indentfirst}
\usepackage{subfigure}
\usepackage{graphicx}
\usepackage[top=1in,bottom=1in,left=1.25in,right=1.25in]{geometry}

\makeatother
\theoremstyle{plain}
\begin{document}

\title{
  \bf  The mean velocity of two-state models of molecular motor
}

\author{Yunxin Zhang}\email[Email: ]{xyz@fudan.edu.cn}
\affiliation{Shanghai Key Laboratory for Contemporary Applied Mathematics,
Centre for Computational System Biology,\\
School of Mathematical Sciences, Fudan University, Shanghai 200433, China.}

\begin{abstract}
\normalsize
The motion of molecular motor is essential to the biophysical functioning of living cells. In principle, this motion can be regraded as a multiple chemical states process. In which, the molecular motor can jump between different chemical states, and in each chemical state, the motor moves forward or backward in a corresponding potential. So, mathematically, the motion of molecular motor can be described by several coupled one-dimensional hopping models or by several coupled Fokker-Planck equations.
To know the basic properties of molecular motor, in this paper, we will give detailed analysis about the simplest cases: in which there are only two chemical states. Actually, many of the existing models, such as the flashing ratchet model, can be regarded as a two-state model. From the explicit expression of the mean velocity, we find that the mean velocity of molecular motor might be nonzero even if the potential in each state is periodic, which means that there is no energy input to the molecular motor in each of the two states. At the same time, the mean velocity might be zero even if there is energy input to the molecular motor. Generally, the velocity of molecular motor depends not only on the potentials (or corresponding forward and backward transition rates) in the two states, but also on the transition rates between the two chemical states.

\end{abstract}

%\date{\today}



\maketitle

\section{Introduction}
Molecular motors are biogenic force generators acting in the nanometer range, and converting chemical energy into mechanical work \cite{Bray2001, Howard2001}, which play essential roles in eukaryotic cells \cite{Badoual2002, Lipowsky2005, Riedel2007, Zhang2009, Howard2009}. In the super family of molecular motors \cite{Vale2003}, the most extensively studied ones are conventional kinesin \cite{Fisher2001, Carter2005, Block2007, Zhang2008, Toprak2009, Guydosh2009, Hyeon2009, Hariharan2009}, cytoplasmic dynein \cite{Samara2006, Toba2006, Gennerich2009, Houdusse2009, Roberts2009, Kardona2009, Serohijos2009}, myosin V \cite{Rosenfeld2004, Purcell2005, Veigel2005, Sakamoto2008, Jackson2009, Fedorov2009}, and ${\rm F_0F_1}-$ATPase \cite{Wang1998, Kazuhiko2000, Nishizaka2004, Adachi2007, Muneyuki2007, Junge2009, Miller2009}. The conventional kinesin can walk hand-over-hand along microtubule about 1 $\mu$m to the plus end direction of the microtubule before its dissociation from the track \cite{Block1990, Yildiz2004, Asbury2003}, with step size 8.2 nm \cite{Schnitzer1997, Coy1999, Fehr2007} and stall force 6$-$8 pN \cite{Guydosh2009, Gennerich2009, Yildiz2008, Block2007, Hackney2005, Taniguchi2005, Carter2005, Nishiyama2002, Schnitzer2000}, which is independent of ATP concentration \cite{Carter2005}. In saturating ATP solution, its zero load velocity is about 700$-$1000 nm/s \cite{Nishiyama2002, Carter2005, Block2003}. Cytoplasmic dynein also can walk hand-over-hand along microtubule with average step size 8.2 nm \cite{Kardona2009, Gennerich2007, Watanabe2007, Mallik2004, Hirakawa2000}, but to the minus end direction \cite{Toba2006}.  Recent experimental data indicate that its stall force is also about 6$-$8 pN \cite{Gennerich2007, Hirakawa2000, Cho2008}, and independent of ATP concentration \cite{Gennerich2007}. To the dynein which is purified from mammalian animals, its maximal velocity is also about 700$-$1000 nm/s \cite{Toba2006, Ross2008, King2000}. Myosin V is also a processive motor but walks along actin filaments with average step size 36 nm, and ATP independent stall force 2$-$3 pN \cite{Cappello2007, Tsygankov2007, Christof2006, Clemen2005, Kolomeisky2003, Uemura2004}. ATPase consists of two portions ${\rm F_0}$ and ${\rm F_1}$ connected to a $\gamma$ shaft. It can use the proton-motive force across the mitochondrial membranes to make ATP from ADP and Pi, and also can use ATP to drive the rotation of the $\gamma$ shaft \cite{Oster2000}. Recent experiments found that there are also many other molecular motors that can move processively, such as kinesin CENP-E \cite{Yardimci2008}, myosin VI \cite{Sweeney2007, Oguchi2008, Bryant2007, Iwaki2009}, myosin VIIa \cite{Udovichenko2002}, myosin IXb \cite{Inoue2002}, myosin XI \cite{Tominaga2003}, and T7 DNA helicase \cite{Kim2002}.

There are many mathematical models to describe the motion of molecular motor, such as Fokker-Planck equation \cite{Risken1989, Zhang20091, Wang2002, Howard2001}, Langevin equation \cite{Gehlen2008}, and master equation \cite{Fisher2001, Nieuwenhuizen2004, Kolomeisky2007, Liepelt2007, Zhang20093}. However, so far, almost all of the explicit formulations of biophysical properties of molecular motor, such as mean velocity \cite{Fisher1999, Howard2001}, effective diffusion constant \cite{Reimann2001, Zhang20092}, and mean first passage time \cite{Pury2003, Kolomeisky2005}, are obtained by employing one-sate models, in which the molecular motor moves along its track in one tilted periodic potential \footnote{The one-state model can be regarded as a simplification of the multi-state model but with an effective potential, Theoretically, this effective potential can be obtained by weighted average of potentials in each state of the multi-state model \cite{Wang2004}.}. One of the basic properties of such models is that the mean velocity of molecular motor does not vanish as long as the input energy is positive.  These models and their corresponding results are valuable to describe the {\it tightly} mechanochemical coupled cases of motor motion. However, recent experimental data indicate the motion of molecular motors, including conventional kinesin \cite{Bieling2008, Endres2006, Seidel2008, Shaevitz2005, Yildiz2008}, cytoplasmic dynein \cite{Gao2006}, myosin II \cite{Nishikawa2008, Masuda2009}, and $\textrm{F}_1$-ATPase \cite{Gerritsma2009} are usually {\it loosely} coupled to ATP hydrolysis, i.e., the input energy might be nonzero even if the mean velocity vanishes. To study these loosely coupled cases, it is necessary to use multi-state models. In fact, the multi-state models have been used by some authors \cite{Lipowsky2000, Lipowsky2003, Zhang20091}. However, it is hard to get meaningful explicit results for the general $N$-state models. Usually the numerical calculations are employed \cite{Chen1999, Wang2003, Wang2004}. 

In this paper, we will give a detailed theoretical analysis to the two-state models. Actually, the two-state models have most of the essential properties of the general multi-state models, and they have been used in many studies \cite{Astumian1997, Parmeggiani1999, Parrondo2002, Reimann20021, Chen1999, Bier1993, Frank1995, Prost1994}. There are two different forms of two-state models: (1) two coupled one-dimensional hopping models, and (2) two coupled one-dimensional Fokker-Planck equations, which is equivalent to two coupled Langevin equations (in fact, it also can be verified that, any one-dimensional hopping model can be well approximated by a one-dimensional Fokker-Planck equation \cite{Zhang2010}). In the following, we will give the explicit formulation of the mean velocity of molecular motor by using two coupled one-dimensional hopping models and two coupled one-dimensional Fokker-Plank equations respectively. From this formulation, the {\it stall force}, i.e., the external load under which the mean velocity vanishes, can be obtained. We find that, the mean velocity, and consequently the stall force depend not only on potentials in the two states (or corresponding forward and backward transition rates), but also on transition rates between the two chemical states. In general, part of the input energy will dissipate into the environment, and so the {\it energy efficiency}, i.e., the ratio of mechanical work done by the molecular motor to the input energy, might be far less than 1. For example, the mean velocity might be zero even if the input energy in each state is nonzero.

The organization of this paper is as follows. In the next section, the two coupled one-dimensional hopping models are discussed, and then in Section {\bf III}, the two coupled Fokker-Planck equations are analyzed. In each models, three special cases are further analyzed: (1) The motor can jump between the two chemical states at only one position. In fact, the properties of this special case are much similar as the usual one state model. At steady state, there is no energy input to the molecular motor during its transition between the two chemical states. (2) The motor can jump between the two states at two positions. This special case has the typical properties of the general cases. (3) One of the two potentials is constant, or all the corresponding transition rates (forward and backward) are equal to each other in one of the two states. This special case corresponds to the flashing ratchet model of molecular motors. Finally, the results are briefly summarized in Section {\bf IV}.

\section{Two coupled one-dimensional hopping models}
\begin{figure}
  \includegraphics[width=400pt]{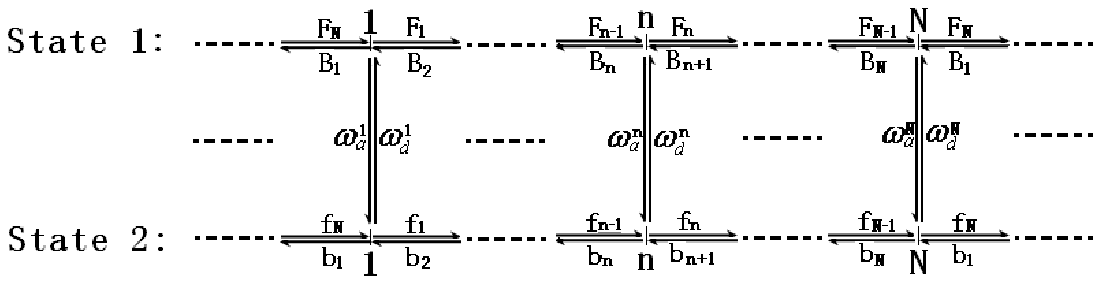}\\
  \caption{Schematic depiction of two coupled one-dimensional hopping models. In which the forward and backward transition rates of molecular motor in state 1 are denoted by $F_n$ and $B_n$, and are denoted by $f_n$ and $b_n$ for molecular motor in state 2, here $1\le n\le N$ with $N$ is the period of the hopping models. The inter-state transition rates at position $n$ are denoted by $\omega_a^n$ (states 1$\to$2) and $\omega_d^n$ (states 2$\to$1). For motor proteins, $\omega_a^n$, $\omega_d^n$ depend on the chemical potentials and concentrations of ATP and ADP. }\label{Fig1}
\end{figure}

The two coupled one-dimensional hopping models are schematically depicted in Fig. \ref{Fig1}. In which, the forward and backward transition rates in state 1 are denoted by $F_n$ ($n\to n+1$) and $B_n$ ($n\to n-1$), the forward and backward transition rates in state 2 are denoted by $f_n$ ($n\to n+1$) and $b_n$ ($n\to n-1$), and the transition rates between the two states at position $n$ are denoted by $\omega^n_a$ (state 1 $\to$ state 2) and $\omega^n_d$ (state 2 $\to$ state 1). Under the assumption of periodicity, we have
\begin{equation}\label{eq1}
\begin{array}{lll}
F_{lN+n}=F_n,\quad &B_{lN+n}=B_n,\quad &\omega_a^{lN+n}=\omega^n_d,\cr
f_{lN+n}=f_n, &b_{lN+n}=b_n, &\omega_d^{lN+n}=\omega^n_d,
\end{array}
\end{equation}
where $l$ is an integer number $l$, $N$ is the period of hopping models. Let $\tilde{P}_n(t)$ be the probability of finding molecular motor at position $n$ of state 1 (denoted by $\textsf{1}_n$) at time $t$, and $\tilde{\rho}_n(t)$ be the probability of finding molecular motor at position $n$ of state 2 (denoted by $\textsf{2}_n$) at time $t$. Then the evolution of probabilities $\tilde{P}_n(t)$ and $\tilde{\rho}_n(t)$ are governed by the following master equations:
\begin{equation}\label{eq2}
\left\{\begin{aligned}
&\frac{d}{dt}\tilde{P}_n(t)=F_{n-1}\tilde{P}_{n-1}(t)-(F_n+B_n)\tilde{P}_n(t)+B_{n+1}\tilde{P}_{n+1}(t)\cr
&\qquad\qquad
-\omega_n^a\tilde{P}_n(t)+\omega_n^d\tilde{\rho}_n(t),\cr
&\frac{d}{dt}\tilde{\rho}_n(t)=f_{n-1}\tilde{\rho}_{n-1}(t)-(f_n+b_n)\tilde{\rho}_n(t)+b_{n+1}\tilde{\rho}_{n+1}(t)\cr
&\qquad\qquad+\omega_n^a\tilde{P}_n(t)-\omega_n^d\tilde{\rho}_n(t),\qquad\qquad
n=0, \pm 1,\pm 2,\cdots.
\end{aligned}\right.
\end{equation}
Let
\begin{equation}\label{eq3}
P_n(t)=\sum_{l=-\infty}^{\infty}\tilde{P}_{lN+n}(t),\qquad
\rho_n(t)=\sum_{l=-\infty}^{\infty}\tilde{\rho}_{lN+n}(t),
\end{equation}
then, at steady state, $P_n$ and $\rho_n$ satisfy \cite{Derrida1983}
\begin{equation}\label{eq4}
\begin{aligned}
&\left\{\begin{aligned}
&F_{n-1}P_{n-1}-(F_n+B_n)P_n+B_{n+1}P_{n+1}-\omega_n^aP_n+\omega_n^d\rho_n=0,\cr
&f_{n-1}\rho_{n-1}-(f_n+b_n)\rho_n+b_{n+1}\rho_{n+1}+\omega_n^aP_n-\omega_n^d\rho_n=0,
\end{aligned}\right.
\end{aligned}
\end{equation}
with $n=1,2,\cdots, N$, and the total flux of probability,
\begin{equation}\label{eq5}
J_{n+\frac12}=(F_nP_n-B_{n+1}P_{n+1})+(f_n\rho_n-b_{n+1}\rho_{n+1}),
\end{equation}
is constant, i.e. $J_{n+\frac12}\equiv J$ for $n=1,2,\cdots, N$.

From the first equation of (\ref{eq4}), one sees that
\begin{equation}\label{eq6}
\rho_n=[(F_n+B_n+\omega^n_a)P_n-F_{n-1}P_{n-1}-B_{n+1}P_{n+1}]/\omega^n_d.
\end{equation}
Substituting (\ref{eq6}) into (\ref{eq5}), one can easily verify
\begin{equation}\label{eq7}
J=A_{n-1}P_{n-1}+C_nP_n+D_{n+1}P_{n+1}+E_{n+2}P_{n+2},
\end{equation}
where
\begin{equation}\label{eq8}
\left\{\begin{aligned} A_n&=-f_{n+1}F_n/\omega^{n+1}_d,\cr
C_n&=[f_n(F_n+B_n+\omega^n_a)]/\omega^n_d+F_n+b_{n+1}F_n/\omega^{n+1}_d,\cr
D_n&=-[B_n+f_{n-1}B_n/\omega^{n-1}_d+b_n(f_n+B_n+\omega^n_a)/\omega^n_d],\cr
E_n&=b_{n-1}B_n/\omega^{n-1}_d.
\end{aligned}\right.
\end{equation}
By (\ref{eq7}) and routine analysis, we obtain
\begin{equation}\label{eq9}
P_i=X_iJ+Y_iP_1+Z_iP_2+W_iP_3,
\end{equation}
where
\begin{equation}\label{eq10}
X_N=\frac{1}{A_N}\quad Y_N=-\frac{C_1}{A_N},\quad
Z_N=-\frac{D_2}{A_N},\quad W_N=-\frac{E_3}{A_N},
\end{equation}
\begin{equation}\label{eq11}
\begin{array}{ll}
X_{N-1}=\frac{1}{A_{N-1}}\left(1-\frac{C_N}{A_N}\right),&\quad
Y_{N-1}=\frac{C_NC_1}{A_{N-1}A_N}-\frac{D_1}{A_{N-1}},\cr
Z_{N-1}=\frac{C_ND_2}{A_{N-1}A_N}-\frac{E_2}{A_{N-1}},&\quad
W_{N-1}=\frac{C_NE_3}{A_{N-1}A_N},
\end{array}
\end{equation}
\begin{equation}\label{eq12}
\left\{\begin{aligned}
&X_{N-2}=\frac{1}{A_{N-2}}-\frac{C_{N-1}}{A_{N-2}A_{N-1}}+\frac{C_{N-1}C_N}{A_{N-2}A_{N-1}A_N}-\frac{D_{N}}{A_{N-2}A_{N}},\cr
&Y_{N-2}=-\frac{E_1}{A_{N-2}}+\frac{C_{N-1}D_1}{A_{N-2}A_{N-1}}+\frac{D_NC_1}{A_{N-2}A_N}-\frac{C_{N-1}C_{N}C_1}{A_{N-2}A_{N-1}A_{N}},\cr
&Z_{N-2}=\frac{C_{N-1}E_2}{A_{N-2}A_{N-1}}+\frac{D_{N}D_2}{A_{N-2}A_{N}}-\frac{C_{N-1}C_ND_2}{A_{N-2}A_{N-1}A_N},\cr
&W_{N-2}=\frac{D_{N}E_3}{A_{N-2}A_{N}}-\frac{C_{N-1}C_NE_3}{A_{N-2}A_{N-1}A_N},\cr
\end{aligned}\right.
\end{equation}
and for $1\le k\le N-3$,
\begin{equation}\label{eq13}
\left\{\begin{aligned}
&X_{k}=\frac{1-(C_{k+1}X_{k+1}+D_{k+2}X_{k+2}+E_{k+3}X_{k+3})}{A_k},\cr
&Y_{k}=-\frac{C_{k+1}Y_{k+1}+D_{k+2}Y_{k+2}+E_{k+3}Y_{k+3}}{A_k},\cr
&Z_{k}=-\frac{C_{k+1}Z_{k+1}+D_{k+2}Z_{k+2}+E_{k+3}Z_{k+3}}{A_k},\cr
&W_{k}=-\frac{C_{k+1}W_{k+1}+D_{k+2}W_{k+2}+E_{k+3}W_{k+3}}{A_k}.
\end{aligned}\right.
\end{equation}
Then, for $i=1,2,3$ in equation (\ref{eq9}), we obtain the following
equations
\begin{equation}\label{eq14}
AP=JX,
\end{equation}
where
\begin{equation}\label{eq15}
 A=\left(\begin{array}{ccc} 1-Y_1 &
-Z_1 & -W_1\cr-Y_2 & 1-Z_2 & -W_2\cr -Y_3 & -Z_3 & 1-W_3
\end{array}\right),\quad P=\left(\begin{array}{c} P_1\cr P_2\cr
P_3\end{array}\right),\quad X=\left(\begin{array}{c} X_1\cr X_2\cr
X_3\end{array}\right).
\end{equation}
So, $P_i=\hat{P}_iJ$ for $i=1,2,3$, with $\hat{P}=(\hat{P}_1, \hat{P}_2, \hat{P}_3)^T$ satisfy $A\hat{P}=X$.
Consequently, $P_i$, for $3<i\le N$, can be obtained by Eq. (\ref{eq9}),
\begin{equation}\label{eq16}
P_i=(X_i+Y_i\hat{P}_1+Z_i\hat{P}_2+W_i\hat{P}_3)J=:\hat{P}_iJ,
\end{equation}
and therefore, $\rho_i$ can be obtained by Eq. (\ref{eq6}),
\begin{equation}\label{eq17}
\begin{aligned}
\rho_i=&[(F_i+B_i+\omega^i_a)P_i-F_{i-1}P_{i-1}-B_{i+1}P_{i+1}]/\omega^i_d,\cr
=&[(F_i+B_i+\omega^i_a)\hat{P}_i-F_{i-1}\hat{P}_{i-1}-B_{i+1}\hat{P}_{i+1}]J/\omega^i_d,\cr
=&:\hat{\rho}_iJ.
\end{aligned}
\end{equation}
The probability flux $J$ in Eqs. (\ref{eq16}) (\ref{eq17}) is determined by the normalization condition
$\sum\limits_{i=1}^N(P_i+\rho_i)=1$,
\begin{equation}\label{eq18}
\begin{aligned}
J=&1/\sum\limits_{i=1}^N(\hat{P}_i+\hat{\rho}_i),\cr
=&1/\sum\limits_{i=1}^N\left[\frac{\omega^i_a+\omega^i_d}{\omega^i_d}+\left(\frac{1}{\omega^i_d}-\frac{1}{\omega^{i+1}_d}\right)F_i+\left(\frac{1}{\omega^i_d}-\frac{1}{\omega^{i-1}_d}\right)B_i\right]\hat{P}_i.
\end{aligned}
\end{equation}
Specially, if $\omega^i_a\equiv \omega_a$ and $\omega^i_d\equiv \omega_d$ are constants, then
\begin{equation}\label{eq19}
\begin{aligned}
J=\frac{\omega_d}{(\omega_a+\omega_d)\sum\limits_{i=1}^N\hat{P}_i}=\frac{\omega_d}{(\omega_a+\omega_d)\sum\limits_{i=1}^N(X_i+Y_i\hat{P}_1+Z_i\hat{P}_2+W_i\hat{P}_3)}.
\end{aligned}
\end{equation}

\subsection{Special case I: $\omega^i_a=\omega^i_d=0\ \textrm{for }1\le i\le N-1$}
\begin{figure}
  \includegraphics[width=400pt]{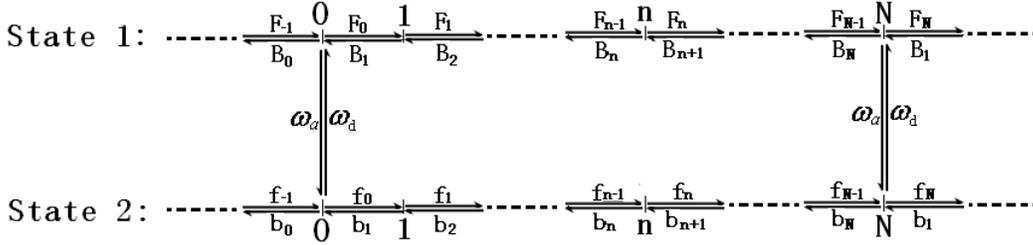}\\
  \caption{Special case {\bf I} of two coupled one-dimensional hopping models. In which $\omega_a^n=\omega_d^n=0$ for $1\le n\le N-1$, and $\omega_a^N=\omega_a$, $\omega_d^N=\omega_d$. For this special case, the mean velocity of molecular motor would be zero if there is no energy input in each of the two states, i.e., $\Delta G_1=\Delta G_2=0$. In fact, at steady state, there is also no energy input during the process that drives the motor from one state to another, since $\omega_aP_N=\omega_d\rho_N$.  }\label{Fig2}
\end{figure}

For convenience, we denote $\omega^N_a, \omega^N_d$ by $\omega_a, \omega_d$ respectively (see Fig. \ref{Fig2}). For this special case, the steady state probabilities $P_n, \rho_n$ satisfy
\begin{equation}\label{eq20}
\left\{\begin{aligned}
&F_NP_N-B_1P_1=F_1P_1-B_2P_2=\cdots=F_{N-1}P_{N-1}-B_NP_N=:J,\cr &
f_N\rho_N-b_1\rho_1=f_1\rho_1-b_2\rho_2=\cdots=f_{N-1}\rho_{N-1}-b_N\rho_N=:j.
\end{aligned}\right.
\end{equation}
It can be readily verified that
\begin{equation}\label{eq21}
P_k=X_kP_N-Y_kJ,
\end{equation}
with
\begin{equation}\label{eq21of1}
\begin{aligned}
X_k=\prod_{i=1}^k\frac{F_{i-1}}{B_i},\quad
Y_k=\frac{1}{F_k}\sum_{i=1}^{k}\prod_{j=i}^{k}\frac{F_j}{B_j}.
\end{aligned}
\end{equation}
Specially,
\begin{equation}\label{eq22}
P_N=\left(\prod_{i=1}^N\frac{F_{i-1}}{B_i}\right)P_N-\left(\frac{1}{F_N}\sum_{i=1}^{N}\prod_{j=i}^{N}\frac{F_j}{B_j}\right)J,
\end{equation}
which implies
\begin{equation}\label{eq23}
P_N=\frac{\frac{1}{F_N}\sum\limits_{i=1}^{N}\prod\limits_{j=i}^{N}\frac{F_j}{B_j}}{\prod\limits_{i=1}^N\frac{F_{i-1}}{B_i}-1}J.
\end{equation}
Combining (\ref{eq21}) (\ref{eq21of1}) and (\ref{eq23}), one finds
\begin{equation}\label{eq24}
\begin{aligned}
P_k=&\left(\frac{\frac{1}{F_N}\left(\sum\limits_{i=1}^{N}\prod\limits_{j=i}^{N}\frac{F_j}{B_j}\right)\left(\prod\limits_{i=1}^k\frac{F_{i-1}}{B_i}\right)}{\prod\limits_{i=1}^N\frac{F_{i-1}}{B_i}-1}
-\frac{1}{F_k}\sum_{i=1}^{k}\prod_{j=i}^{k}\frac{F_j}{B_j}\right)J,\cr
=&\frac{\left(\sum\limits_{i=1}^{N}\prod\limits_{j=i}^{N}\frac{F_j}{B_j}\right)\left(\prod\limits_{i=1}^{k}\frac{F_{i}}{B_i}\right)-\left(\sum\limits_{i=1}^{k}\prod\limits_{j=i}^{k}\frac{F_j}{B_j}\right)\left(\prod\limits_{i=1}^N\frac{F_{i-1}}{B_i}-1\right)}{\prod\limits_{i=1}^N\frac{F_{i-1}}{B_i}-1}\frac{J}{F_k}.
\end{aligned}
\end{equation}
Using the periodic conditions (\ref{eq1}), one can verify that
\begin{equation}\label{eq25}
\begin{aligned}
P_k
=\frac{\frac{1}{F_k}\sum\limits_{i=k+1}^{N+k}\prod\limits_{j=i}^{N+k}\frac{F_j}{B_j}}{\prod\limits_{i=1}^N\frac{F_{i}}{B_i}-1}J.
\end{aligned}
\end{equation}
Using the same method, the probability $\rho_k$ can be obtained
\begin{equation}\label{eq26}
\begin{aligned}
\rho_k
=\frac{\frac{1}{f_k}\sum\limits_{i=k+1}^{N+k}\prod\limits_{j=i}^{N+k}\frac{f_j}{b_j}}{\prod\limits_{i=1}^N\frac{f_{i}}{b_i}-1}j.
\end{aligned}
\end{equation}
At steady state, $\omega_aP_N=\omega_d\rho_N$, which implies
\begin{equation}\label{eq27}
j=\frac{\frac{1}{F_N}\sum\limits_{i=1}^{N}\prod\limits_{j=i}^{N}\frac{F_j}{B_j}\left/\left(\prod\limits_{i=1}^N\frac{F_{i}}{B_i}-1\right)\right.}
{\frac{1}{f_N}\sum\limits_{i=1}^{N}\prod\limits_{j=i}^{N}\frac{f_j}{b_j}\left/\left(\prod\limits_{i=1}^N\frac{f_{i}}{b_i}-1\right)\right.}\frac{\omega_a}{\omega_d}J=:\Xi
J.
\end{equation}
Therefore,
\begin{equation}\label{eq28}
\rho_k
=\frac{\frac{1}{f_k}\sum\limits_{i=k+1}^{N+k}\prod\limits_{j=i}^{N+k}\frac{f_j}{b_j}}{\prod\limits_{i=1}^N\frac{F_{i}}{B_i}-1}\Theta
J,\qquad\textrm{where}\ \
\Theta=\frac{\frac{\omega_a}{F_N}\sum\limits_{i=1}^{N}\prod\limits_{j=i}^{N}\frac{F_j}{B_j}}{\frac{\omega_d}{f_N}\sum\limits_{i=1}^{N}\prod\limits_{j=i}^{N}\frac{f_j}{b_j}}.
\end{equation}
Since $P_k, \rho_k$ satisfy $\sum\limits_{k+1}^N(P_k+\rho_k)=1$,
from (\ref{eq25}, \ref{eq28}), the probability flux $J$ can be obtained as follows
\begin{equation}\label{eq29}
\begin{aligned}
J=\frac{\phi\left(\prod\limits_{i=1}^N\frac{F_{i}}{B_i}-1\right)}{\phi\Psi+\Phi\psi}.
\end{aligned}
\end{equation}
where
\begin{equation}\label{eq29of1}
\begin{array}{ll}
\phi=\frac{\omega_d}{f_N}\sum\limits_{i=1}^{N}\prod\limits_{j=i}^{N}\frac{f_j}{b_j},\qquad
&\psi=\sum\limits_{k=1}^N\left(\frac{1}{f_k}\sum\limits_{i=k+1}^{N+k}\prod\limits_{j=i}^{N+k}\frac{f_j}{b_j}\right),\cr
\Phi=\frac{\omega_a}{F_N}\sum\limits_{i=1}^{N}\prod\limits_{j=i}^{N}\frac{F_j}{B_j},
&\Psi=\sum\limits_{k=1}^N\left(\frac{1}{F_k}\sum\limits_{i=k+1}^{N+k}\prod\limits_{j=i}^{N+k}\frac{F_j}{B_j}\right).
\end{array}
\end{equation}
So the total flux of this system is
\begin{equation}\label{eq30}
\begin{aligned}
J+j=(1+\Xi
)J=\left(1+\frac{\prod\limits_{i=1}^N\frac{f_{i}}{b_i}-1}{\prod\limits_{i=1}^N\frac{F_{i}}{B_i}-1}\Theta\right)J
=\frac{\phi\left(\prod\limits_{i=1}^N\frac{F_{i}}{B_i}-1\right)+\Phi\left(\prod\limits_{i=1}^N\frac{f_{i}}{b_i}-1\right)}{\phi\Psi+\Phi\psi}.
\end{aligned}
\end{equation}
Combining (\ref{eq25}) (\ref{eq28}) and (\ref{eq29}), the probabilities $P_k$ and $\rho_k$ can be obtained as follows
\begin{equation}\label{eq31}
\begin{aligned}
P_k=\frac{\phi}{\phi\Psi+\Phi\psi}\frac{1}{F_k}\sum\limits_{i=k+1}^{N+k}\prod\limits_{j=i}^{N+k}\frac{F_j}{B_j},\quad
\rho_k=\frac{\Phi}{\phi\Psi+\Phi\psi}\frac{1}{f_k}\sum\limits_{i=k+1}^{N+k}\prod\limits_{j=i}^{N+k}\frac{f_j}{b_j}.
\end{aligned}
\end{equation}
By (\ref{eq30}), one easily sees that, if $\prod\limits_{i=1}^N\frac{f_{i}}{b_i}=\prod\limits_{i=1}^N\frac{F_{i}}{B_i}=1$,
then $J=j=0$, consequently the total probability flux $J+j=0$. In other words, for this special case, if there is no energy input to the molecular motor in each state, then the mean velocity would be zero. But the reverse does not hold. Note, the potential changes in one period of state 1 and state 2 are $\Delta G_1=k_BT\ln\left(\prod_{i=1}^N\frac{F_i}{B_i}\right)$ and $\Delta G_2=k_BT\ln\left(\prod_{i=1}^N\frac{f_i}{b_i}\right)$ respectively \cite{Qian1997, Fisher2001}.

\subsection{Special case II:  $\omega^i_a=\omega^i_d=0\ \textrm{for}\ i\ne M,N$}
\begin{figure}
  \includegraphics[width=400pt]{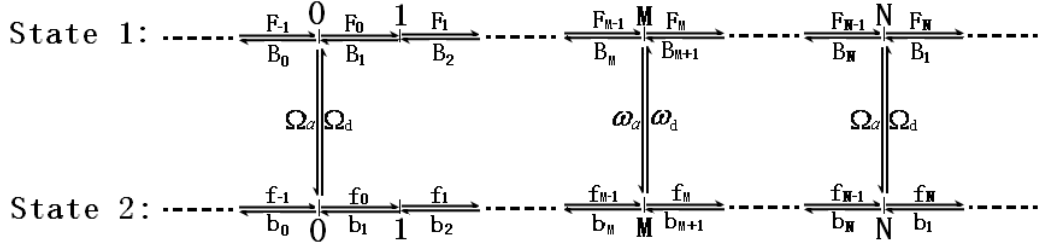}\\
  \caption{Special case {\bf II} of two coupled one-dimensional hopping models. In which $\omega_a^n=\omega_d^n=0$ for $n\ne M, N$, and $\omega_a^M=\omega_a$, $\omega_d^M=\omega_d$, $\omega_a^N=\Omega_a$, $\omega_d^N=\Omega_d$. For this special case, the mean velocity of molecular motors might not be zero even if there is no energy input in each of the two states, i.e., $\Delta G_1=\Delta G_2=0$. Since there usually exists energy input to molecular motor during its jump from one state to another unless $\omega_aP_M=\omega_d\rho_M$ and $\Omega_aP_N=\Omega_d\rho_N$. }\label{Fig3}
\end{figure}

For  convenience, we denote $\omega^N_a, \omega^N_d$ by $\Omega_a, \Omega_d$, and  $\omega^M_a, \omega^M_d$ by $\omega_a, \omega_d$ (see Fig. \ref{Fig3}). At steady state, $P_k, \rho_k$ satisfy
\begin{equation}\label{eq32}
\left\{\begin{aligned} &F_NP_N-B_1P_1=F_1P_1-B_2P_2=\cdots
=F_{M-1}P_{M-1}-B_MP_M=:J_1,\cr
&F_MP_M-B_{M+1}P_{M+1}
=\cdots=F_{N-1}P_{N-1}-B_NP_N=:J_2,\cr
&f_N\rho_N-b_1\rho_1=f_1\rho_1-b_2\rho_2=\cdots
=f_{M-1}\rho_{M-1}-b_M\rho_M=:j_1,\cr
&f_M\rho_M-b_{M+1}\rho_{M+1}
=\cdots=f_{N-1}\rho_{N-1}-b_N\rho_N=:j_2,\cr
&\omega_aP_M+\Omega_aP_N=\omega_d\rho_M+\Omega_d\rho_N,\cr
&J_2=J_1+\Omega_aP_N-\Omega_d\rho_N,\cr
&j_2=j_1-\Omega_aP_N+\Omega_d\rho_N,\cr
&\sum\limits_{k=1}^N(P_k+\rho_k)=1.
\end{aligned}\right.
\end{equation}
From the first equation in (\ref{eq32}), one can easily get
\begin{equation}\label{eq33}
\begin{aligned}
P_k=&\left(\prod_{i=1}^k\frac{F_{i-1}}{B_i}\right)P_N-\frac{1}{F_k}\left(\sum_{i=1}^{k}\prod_{j=i}^{k}\frac{F_j}{B_j}\right)J_1,
\quad 1\le k\le M.
\end{aligned}
\end{equation}
At the same time, from the second equation in (\ref{eq32}),
\begin{equation}\label{eq34}
\begin{aligned}
P_k=&\left(\prod_{i=M+1}^k\frac{F_{i-1}}{B_i}\right)P_M-\frac{1}{F_k}\left(\sum_{i=M+1}^{k}\prod_{j=i}^{k}\frac{F_j}{B_j}\right)J_2\cr
=&\left(\prod_{i=M+1}^k\frac{F_{i-1}}{B_i}\right)\left[\left(\prod_{i=1}^M\frac{F_{i-1}}{B_i}\right)P_N-\frac{1}{F_M}\left(\sum_{i=1}^{M}\prod_{j=i}^{M}\frac{F_j}{B_j}\right)J_1\right]\cr
&-\frac{1}{F_k}\left(\sum_{i=M+1}^{k}\prod_{j=i}^{k}\frac{F_j}{B_j}\right)(J_1+\Omega_aP_N-\Omega_d\rho_N)\cr
=&\left[\prod_{i=1}^k\frac{F_{i-1}}{B_i}-\frac{\Omega_a}{F_k}\left(\sum_{i=M+1}^{k}\prod_{j=i}^{k}\frac{F_j}{B_j}\right)\right]P_N\cr
&+\frac{\Omega_d}{F_k}\left(\sum_{i=M+1}^{k}\prod_{j=i}^{k}\frac{F_j}{B_j}\right)\rho_N
-\frac{1}{F_k}\left(\sum_{i=1}^{k}\prod_{j=i}^{k}\frac{F_j}{B_j}\right)J_1\cr
=&:(R_k-\Omega_aS_k)P_N+\Omega_dS_k\rho_N-T_kJ_1, \quad \textrm{for } M+1\le k\le N.
\end{aligned}
\end{equation}
In particular, $P_N=(R_N-\Omega_aS_N)P_N+\Omega_dS_N\rho_N-T_NJ_1$,
which gives
\begin{equation}\label{eq35}
J_1=\frac{(R_N-\Omega_aS_N-1)P_N+\Omega_dS_N\rho_N}{T_N}.
\end{equation}
Substituting (\ref{eq35}) into (\ref{eq33}) (\ref{eq34}), we obtain
\begin{equation}\label{eq36}
P_k=G_kP_N+H_k\rho_N,
\end{equation}
where
\begin{equation}\label{eq37}
\begin{aligned}
G_k=&\left\{\begin{array}{ll}
R_k-\frac{T_k}{T_N}(R_N-\Omega_aS_N-1),\qquad &1\le k\le M,\cr
R_k-\Omega_aS_k-\frac{T_k}{T_N}(R_N-\Omega_aS_N-1), &M+1\le k\le N,
\end{array}\right.\cr
H_k=&\left\{\begin{array}{ll}
-\frac{T_k}{T_N}\Omega_dS_N,\qquad\qquad\qquad &1\le k\le M,\cr
-\frac{T_k}{T_N}\Omega_dS_N+\Omega_dS_k, & M+1\le k\le N.
\end{array}\right.
\end{aligned}
\end{equation}
Similarly,
\begin{equation}\label{eq38}
\rho_k=g_k\rho_N+h_kP_N,
\end{equation}
where $g_k, h_k$, and the corresponding $r_k, s_k, t_k$ in expressions of $g_k, h_k$, can be obtained by replacing $F_j, B_j, \Omega_a, \Omega_d$ in the expressions of $R_k, S_k, T_k, G_k, H_k$ with $f_j, b_j, \Omega_d, \Omega_a$ respectively. Combining (\ref{eq36}) (\ref{eq38}) and the fifth equality in (\ref{eq32}), we have
\begin{equation}\label{eq39}
\omega_a(G_MP_N+H_M\rho_N)+\Omega_aP_N=\omega_d(g_M\rho_N+h_MP_N)+\Omega_d\rho_N,
\end{equation}
i.e.
\begin{equation}\label{eq39of1}
(\Omega_a+\omega_aG_M-\omega_dh_M)P_N=(\Omega_d+\omega_dg_M-\omega_aH_M)\rho_N.
\end{equation}
So
\begin{equation}\label{eq40}
\rho_N=\frac{\Omega_a+\omega_aG_M-\omega_dh_M}{\Omega_d+\omega_dg_M-\omega_aH_M}P_N=:\frac{U}{V}P_N.
\end{equation}
From (\ref{eq36}) (\ref{eq38}) and (\ref{eq40}), one finds
\begin{equation}\label{eq41}
P_k=\left(G_k+\frac{U}{V}H_k\right)P_N,\qquad
\rho_k=\left(h_k+\frac{U}{V}g_k\right)P_N.
\end{equation}
In view of the last equation in (\ref{eq32}), one gets
\begin{equation}\label{eq41of1}
\left[\sum_{k=1}^N\left((G_k+h_k)+\frac{U}{V}(g_k+H_k)\right)\right]P_N=1,
\end{equation}
which implies
\begin{equation}\label{eq42}
P_N=\frac{1}{\sum\limits_{k=1}^N\left[(G_k+h_k)+\frac{U}{V}(g_k+H_k)\right]}.
\end{equation}
By (\ref{eq35}) (\ref{eq40}) (\ref{eq42}), we have
\begin{equation}\label{eq43}
J_1=\frac{(R_N-\Omega_aS_N-1)+\Omega_dS_N\frac{U}{V}}{T_N\sum\limits_{k=1}^N\left[(G_k+h_k)+\frac{U}{V}(g_k+H_k)\right]}.
\end{equation}
Similarly, one can verify that
\begin{equation}\label{eq44}
j_1=\frac{(r_N-\Omega_ds_N-1)\frac{U}{V}+\Omega_as_N}{t_N\sum\limits_{k=1}^N\left[(G_k+h_k)+\frac{U}{V}(g_k+H_k)\right]}.
\end{equation}
Therefore, the total flux of this special case is
\begin{equation}\label{eq45}
J_1+j_1=\frac{\left((R_N-\Omega_aS_N-1)+\Omega_dS_N\frac{U}{V}\right)t_N+\left((r_N-\Omega_ds_N-1)\frac{U}{V}+\Omega_as_N\right)T_N}{T_Nt_N\sum\limits_{k=1}^N\left[(G_k+h_k)+\frac{U}{V}(g_k+H_k)\right]}.
\end{equation}
More specially, if $R_N=r_N=1$, then the total probability flux is
\begin{equation}\label{eq46}
\begin{aligned}
J_1+j_1=&\frac{1}{\sum\limits_{k=1}^N\left[(G_k+h_k)+\frac{U}{V}(g_k+H_k)\right]}\left(\frac{s_N}{t_N}-\frac{S_N}{T_N}\right)\left(\Omega_a-\frac{U}{V}\Omega_d\right),\cr
=&\frac{\Omega_a\omega_dr_M-\omega_a\Omega_dR_M}{\sum\limits_{k=1}^N\left[(G_k+h_k)V+(g_k+H_k)U\right]}\left(\frac{s_N}{t_N}-\frac{S_N}{T_N}\right),
\end{aligned}
\end{equation}
where
\begin{equation}\label{eq46of1}
\begin{aligned}
U=&\Omega_a+\omega_aR_M+\omega_a\Omega_aS_NT_M/T_N+\omega_d\Omega_as_Nt_M/t_N>0,\cr
V=&\Omega_d+\omega_dr_M+\omega_d\Omega_ds_Nt_M/t_N+\omega_a\Omega_dS_NT_M/T_N>0.
\end{aligned}
\end{equation}
So the direction of probability flux is determined by the sign of $\left(\frac{s_N}{t_N}-\frac{S_N}{T_N}\right)$ and $\left(\Omega_a\omega_dr_M-\omega_a\Omega_dR_M\right)$.
One can see that, $R_N=r_N=1$, i.e., $\Delta G_1=\Delta G_2=0$, does not read the mean velocity vanishes.

To better understand the inter-state transition rates dependence of the total probability flux, we assume that
$$
(\Omega_a, \Omega_d, \omega_a, \omega_d)=\lambda(\tilde\Omega_a, \tilde\Omega_d, \tilde\omega_a, \tilde\omega_d).
$$
It can be verified that the total probability flux $J:=J_1+j_1$ in (\ref{eq46}) increases monotonically with parameter $\lambda$. If $\lambda=0$ then $J=0$. If $\lambda\to \infty$, them $J$ tends to
\begin{equation}\label{eq48}
\begin{aligned}
\frac{\tilde\Omega_a\tilde\omega_dr_M-\tilde\omega_a\tilde\Omega_dR_M}{*}\left(\frac{s_N}{t_N}-\frac{S_N}{T_N}\right),
\end{aligned}
\end{equation}
where
$$
\begin{aligned}
*=&\sum\limits_{k=1}^N\left[(\frac{\tilde\omega_d\tilde\Omega_ds_Nt_M}{t_N}+\frac{\tilde\omega_a\tilde\Omega_dS_NT_M}{T_N})R_k+(\frac{\tilde\omega_a\tilde\Omega_aS_NT_M}{T_N}+\frac{\tilde\omega_d\tilde\Omega_as_Nt_M}{t_N})r_k\right]\cr
&+(\tilde\Omega_a\tilde\omega_dr_M-\tilde\omega_a\tilde\Omega_dR_M)\left[\sum\limits_{k=1}^N\left(\frac{T_kS_N}{T_N}-\frac{t_ks_N}{t_N}\right)
+\sum\limits_{k=M+1}^N(s_k-S_k)\right]\cr
=&\left(\frac{\tilde\omega_ds_Nt_M}{t_N}+\frac{\tilde\omega_aS_NT_M}{T_N}\right)\sum\limits_{k=1}^N(\tilde\Omega_dR_k+\tilde\Omega_ar_k)\cr
&+(\tilde\Omega_a\tilde\omega_dr_M-\tilde\omega_a\tilde\Omega_dR_M)\left[\sum\limits_{k=1}^N\left(\frac{T_kS_N}{T_N}-\frac{t_ks_N}{t_N}\right)
+\sum\limits_{k=M+1}^N(s_k-S_k)\right].
\end{aligned}
$$

\subsection{Special case III:  $\omega^i_a=\omega^i_d=0\ \textrm{for}\ i\ne M,N$, and $f_i=b_i\equiv f$ \ \textrm{for}\ $1\le i\le N$}
As pointed out in the Introduction, the flashing ratchet model can be regarded as one example of this special case. For this more special case, we have
\begin{equation}\label{eq49}
\begin{aligned}
r_k\equiv 1,\qquad t_k=\frac{k}{f},\quad\textrm{for }\ 1\le k\le N,
\end{aligned}
\end{equation}
and $s_k=(k-M)/f$ for $M+1\le k\le N$. It can be easily verified that
\begin{equation}\label{eq50}
\begin{aligned}
g_k=&\left\{\begin{array}{ll} 1+\frac{(N-M)k}{Nf}\Omega_d,\qquad
&1\le k\le M,\cr 1+\frac{M(N-k)}{Nf}\Omega_d, &M+1\le k\le N,
\end{array}\right.\cr
h_k=&\left\{\begin{array}{ll} -\frac{(N-M)k}{Nf}\Omega_a,\qquad\quad
&1\le k\le M,\cr -\frac{M(N-k)}{Nf}\Omega_a, & M+1\le k\le N.
\end{array}\right.
\end{aligned}
\end{equation}
So
\begin{equation}\label{eq51}
\begin{aligned}
U=&\Omega_a+\omega_aG_M-\omega_dh_M\cr
=&\Omega_a+\omega_a\left(R_M+\frac{T_M}{T_N}S_N\Omega_a-\frac{T_M}{T_N}(R_N-1)\right)+\frac{M(N-M)}{Nf}\omega_d\Omega_a,\cr
V=&\Omega_d+\omega_dg_M-\omega_aH_M\cr
=&\Omega_d+\omega_d\left(1+\frac{M(N-M)}{Nf}\Omega_d\right)+\frac{T_M}{T_N}S_N\omega_a\Omega_d.
\end{aligned}
\end{equation}
Moreover, if $R_N=1$ then the total probability flux is
\begin{equation}\label{eq52}
\begin{aligned}
J_1+j_1=&\frac{\Omega_a\omega_d-\omega_a\Omega_dR_M}{\Delta}\left(\frac{N-M}{N}-\frac{S_N}{T_N}\right),
\end{aligned}
\end{equation}
where
$$
\begin{aligned}
\Delta=&\sum\limits_{k=1}^M\left(VR_k+Ur_k\right)+(\Omega_a\omega_dr_M-\omega_a\Omega_dR_M)\left[\sum\limits_{k=1}^M\left(\frac{T_kS_N}{T_N}-\frac{t_ks_N}{t_N}\right)
+\sum\limits_{k=M+1}^N(s_k-S_k)\right],\cr
=&(\Omega_a\omega_d-\omega_a\Omega_dR_M)
\left[\frac{S_N}{T_N}\sum\limits_{k=1}^MT_k-\sum\limits_{k=M+1}^NS_k+\frac{(N-M)[(N-M)(N+M+1)-MN]}{2Nf}\right]\cr
&+MU+V\sum\limits_{k=1}^MR_k.
\end{aligned}
$$

\section{Two coupled Fokker-Planck equations}
The general two coupled Fokker-Planck equations are as follows
\begin{equation}\label{eq53}
\left\{\begin{aligned}
\partial_t\tilde P=&D\partial_x(\beta
V_1'\tilde P+\partial_x\tilde P)+\omega_d(x)\tilde
\rho-\omega_a(x)\tilde P,\cr
\partial_t\tilde \rho=&D\partial_x(\beta
V_2'\tilde \rho+\partial_x\tilde \rho)-\omega_d(x)\tilde
\rho+\omega_a(x)\tilde P,
\end{aligned}\right.\quad -\infty\le x\le +\infty,
\end{equation}
where $D$ is free diffusion constant, $\beta=1/k_BT$ with $k_B$
is Boltzmann constant, and $T$ is absolute temperature, $P(x,t)$ and
$\rho(x,t)$ are probability densities of finding molecular motor at position $x$
at time $t$ and in states 1 and 2 respectively, $V_1, V_2$ are
(tilted) periodic potentials with period $L$. $\omega_a(x),
\omega_d(x)$ are transition rates between states 1 and 2 at
position $x$ \footnote{For motor proteins, $\omega_a(x), \omega_d(x)$ depend on the standard chemical potentials and concentrations of ATP, ADP and ionic phosphate Pi \cite{Howard2001, Parrondo2002}.}.
Similar as in \cite{Zhang20092}, let
\begin{equation}\label{eq55}
P(x, t)=\sum_{k=-\infty}^{+\infty}P(x+kL, t),\qquad \rho(x,
t)=\sum_{k=-\infty}^{+\infty}\rho(x+kL, t),
\end{equation}
then $P(x, t), \rho(x, t)$ satisfy
\begin{equation}\label{eq56}
\left\{\begin{aligned}
\partial_t P=&D\partial_x(\beta
V_1' P+\partial_x P)+\omega_d(x) \rho-\omega_a(x) P,\cr
\partial_t \rho=&D\partial_x(\beta
V_2' \rho+\partial_x \rho)-\omega_d(x) \rho+\omega_a(x) P,
\end{aligned}\right.\quad 0\le x\le L.
\end{equation}

The steady state solution of (\ref{eq56}) can be obtained under
the following constraints:
\begin{equation}\label{eq57}
\begin{aligned}
P(0)=P(L),\quad \rho(0)=\rho(L),\quad \int_0^L(P+\rho)dx=1,\quad
\int_0^L\omega_d\rho dx=\int_0^L\omega_aP dx.
\end{aligned}
\end{equation}
The corresponding probability flux is
\begin{equation}\label{eq58}
\begin{aligned}
J=-D\left(\beta V_1' P+\partial_x P+\beta V_2' \rho+\partial_x
\rho\right),
\end{aligned}
\end{equation}
and the mean velocity of molecular motor is $V=\int_{0}^{L}Jdx=-\beta D\int_{0}^{L}(V_1' P+V_2' \rho)dx=JL$.
If $\omega_a(x), \omega_d(x)$ are constants, Eq. (\ref{eq56}) had been
discussed by Y.-D. Chen \cite{Chen1999}, and it can be solved numerically using
the similar method as the one used in WPE method \cite{Wang2003, Wang2004}.

\subsection{Special case I: $\omega_a(x)=\omega_d(x)\equiv 0$ for $0<x<L$}
For this special case, the steady state probability densities $P(x), \rho(x)$ of finding molecular motor at position $x$ are governed by the following equations
\begin{equation}\label{eq59}
\left\{\begin{aligned}
&D\partial_x(\beta
V_1' P+\partial_x P)=0,\cr
&D\partial_x(\beta
V_2' \rho+\partial_x \rho)=0,
\end{aligned}\right.\quad 0< x< L.
\end{equation}
Meanwhile, $P(x), \rho(x)$ satisfy the following boundary conditions and normalization constraint:
\begin{equation}\label{eq60}
P(0)=P(L),\quad\rho(0)=\rho(L),\quad\omega_aP(0)=\omega_d\rho(0),\quad\int_0^L(P+\rho)dx=1,
\end{equation}
where $\omega_a=\omega_a(L), \omega_d=\omega_d(L)$. The probability fluxes in the two states are
\begin{equation}\label{eq61}
J=-D(\beta
V_1' P+\partial_x P),\qquad j=-D(\beta
V_2' \rho+\partial_x \rho).
\end{equation}
So Eqs. (\ref{eq59}) can be reformulated as
\begin{equation}\label{eq62}
\left\{\begin{aligned}
&\beta V_1' P+\partial_x P=-J/D,\cr
&\beta
V_2' \rho+\partial_x \rho=-j/D,
\end{aligned}\right.\quad 0< x< L.
\end{equation}
The general solutions of (\ref{eq62}) are
$$
P(x)=\left(-\frac{J}{D}\int_0^xe^{\beta V_1(y)}dy+C_1\right)e^{-\beta V_1(x)},\quad
\rho(x)=\left(-\frac{j}{D}\int_0^xe^{\beta V_2(y)}dy+C_2\right)e^{-\beta V_2(x)},
$$
where the constants $C_1, C_2$ can be determined by the periodic boundary conditions $P(0)=P(L), \rho(0)=\rho(L)$:
$$
C_1=\frac{\frac{J}{D}\int_0^Le^{\beta V_1(y)}dy}{1-e^{-\beta\Delta V_1}},\qquad
C_2=\frac{\frac{J}{D}\int_0^Le^{\beta V_2(y)}dy}{1-e^{-\beta\Delta V_2}},
$$
with $\Delta V_i=V_i(0)-V_i(L)$. Therefore
\begin{equation}\label{eq63}
P(x)=\frac{\frac{J}{D}\int_x^{x+L}e^{\beta [V_1(y)-V_1(x)]}dy}{1-e^{-\beta\Delta V_1}},\qquad
\rho(x)=\frac{\frac{j}{D}\int_x^{x+L}e^{\beta [V_2(y)-V_2(x)]}dy}{1-e^{-\beta\Delta V_2}}.
\end{equation}
From $\omega_aP(0)=\omega_d\rho(0)$, one sees
$$
\omega_a\frac{\frac{J}{D}\int_0^{L}e^{\beta [V_1(y)-V_1(0)]}dy}{1-e^{-\beta\Delta V_1}}=
\omega_d\frac{\frac{j}{D}\int_0^{L}e^{\beta [V_2(y)-V_2(0)]}dy}{1-e^{-\beta\Delta V_2}},
$$
so
\begin{equation}\label{eq64}
j=\frac{\omega_a\left(e^{\beta V_2(0)}-(e^{\beta V_2(L)}\right)\int_0^{L}e^{\beta V_1(y)}dy}{\omega_d\left(e^{\beta V_1(0)}-(e^{\beta V_1(L)}\right)\int_0^{L}e^{\beta V_2(y)}dy}J.
\end{equation}
From (\ref{eq63}) (\ref{eq64}) and the normalization condition $\int_0^L(P+\rho)dx=1$, one can easily get
\begin{equation}\label{eq65}
\begin{aligned}
J=&\frac{\omega_dD\left(e^{\beta V_1(0)}-e^{\beta V_1(L)}\right)\int_0^{L}e^{\beta V_2(y)}dy}{\star},\cr
j=&\frac{\omega_aD\left(e^{\beta V_2(0)}-e^{\beta V_2(L)}\right)\int_0^{L}e^{\beta V_1(y)}dy}{\star},
\end{aligned}
\end{equation}
where
$$
\begin{aligned}
\star=&\omega_de^{\beta V_1(0)}\left(\int_0^{L}e^{\beta V_2(y)}dy\right)\left[\int_0^{L}e^{-\beta V_1(x)}\left(\int_x^{x+L}e^{\beta V_1(y)}dy\right)dx\right]\cr
&+\omega_ae^{\beta V_2(0)}\left(\int_0^{L}e^{\beta V_1(y)}dy\right)\left[\int_0^{L}e^{-\beta V_2(x)}\left(\int_x^{x+L}e^{\beta V_2(y)}dy\right)dx\right].
\end{aligned}
$$
It can be easily found that, for this special case, the total probability flux $J+j=0$ if potentials $V_1, V_2$ are periodic, i.e., $\Delta V_1=\Delta V_2=0$. From (\ref{eq30}) and (\ref{eq65}), one sees that, the properties of this special case are similar as those of the special case {\bf I} of the two coupled one-dimensional hopping models \cite{Zhang2010}.

\subsection{Special case II: $\omega_a(x)=\omega_d(x)\equiv 0$ for $x\ne a, L$}

For this special case, the governing equations of the steady state probability densities $P(x), \rho(x)$ are
\begin{equation}\label{eq66}
\begin{aligned}
&\left\{\begin{aligned}
&D\partial_x(\beta
V_1' P_1+\partial_x P_1)=0,\cr
&D\partial_x(\beta
V_2' \rho_1+\partial_x \rho_1)=0,
\end{aligned}\right.\quad 0< x< a,\cr
&\left\{\begin{aligned}
&D\partial_x(\beta
V_1' P_2+\partial_x P_2)=0,\cr
&D\partial_x(\beta
V_2' \rho_2+\partial_x \rho_2)=0,
\end{aligned}\right.\quad a< x< L,
\end{aligned}
\end{equation}
with the following constraints
\begin{equation}\label{eq67}
\begin{aligned}
&P_1(0)=P_2(L),\quad P_1(a)=P_2(a),\cr
&\rho_1(0)=\rho_2(L),\quad \rho_1(a)=\rho_2(a),\cr
&J_1=J_2+\omega_aP(a)-\omega_d\rho(a),\cr
&j_1=j_2-\omega_aP(a)+\omega_d\rho(a),\cr
&\omega_aP(a)+\Omega_aP(L)=\omega_d\rho(a)+\Omega_d\rho(L),\cr
&\int_0^a(P_1+\rho_1)dx+\int_a^L(P_2+\rho_2)dx=1,
\end{aligned}
\end{equation}
where $\omega_a=\omega_a(a), \omega_d=\omega_d(a)$, $\Omega_a=\omega_a(L), \Omega_d=\omega_d(L)$, and
$$
J_i=-D(\beta
V_1' P_i+\partial_x P_i)\qquad j_i=D-(\beta
V_2' \rho_i+\partial_x \rho_i),\qquad\textrm{for } i=1, 2,
$$
are probability fluxes in the two states.

The general solutions of (\ref{eq66}) can be written as follows
\begin{equation}\label{eq68}
\begin{aligned}
P_i(x)=-F_i(x)J_i+G_i(x)C_i,\qquad \rho_i(x)=-f_i(x)j_i+g_i(x)c_i,\qquad i=1, 2,
\end{aligned}
\end{equation}
where
\begin{equation}\label{eq69}
\begin{array}{lll}
F_1(x)=\frac{1}{D}e^{-\beta V_1(x)}\int_0^xe^{\beta V_1(y)}dy,\quad &G_1(x)=e^{-\beta V_1(x)},\quad &0\le x\le a,\cr
F_2(x)=\frac{1}{D}e^{-\beta V_1(x)}\int_a^xe^{\beta V_1(y)}dy,\quad &G_2(x)=e^{-\beta V_1(x)},&a\le x\le L,\cr
f_1(x)=\frac{1}{D}e^{-\beta V_2(x)}\int_0^xe^{\beta V_2(y)}dy,\quad &g_1(x)=e^{-\beta V_2(x)},&0\le x\le a,\cr
f_2(x)=\frac{1}{D}e^{-\beta V_2(x)}\int_a^xe^{\beta V_2(y)}dy,\quad &g_2(x)=e^{-\beta V_2(x)},&a\le x\le L.
\end{array}
\end{equation}
From (\ref{eq67}) (\ref{eq68}), one can verify that $J_i, j_i$ and $C_i, c_i$ satisfy the following equations
\begin{equation}\label{eq70}
\begin{aligned}
&G_1(0)C_1=-F_2(L)J_2+G_2(L)C_2,\cr
&-F_1(a)J_1+G_1(a)C_1=G_2(a)C_2,\cr
&g_1(0)c_1=-f_2(L)j_2+g_2(L)c_2,\cr
&-f_1(a)j_1+g_1(a)c_1=g_2(a)c_2,\cr
&J_1=J_2+\omega_aG_2(a)C_2-\omega_dg_2(a)c_2,\cr
&J_1+j_1=J_2+j_2,\cr
&\omega_aG_2(a)C_2+\Omega_aG_1(0)C_1=\omega_dg(a)c_2+\Omega_dg_1(0)c_1,\cr
&-\left(\int_0^aF_1dx\right)J_1+\left(\int_0^aG_1dx\right)C_1-\left(\int_a^LF_2dx\right)J_2+\left(\int_a^LG_2dx\right)C_2,\cr
&-\left(\int_0^af_1dx\right)j_1+\left(\int_0^ag_1dx\right)c_1-\left(\int_a^Lf_2dx\right)j_2+\left(\int_a^Lg_2dx\right)c_2=1.
\end{aligned}
\end{equation}
For the sake of convenience, we rewrite equations in (\ref{eq70}) as $AX=B$, with $X=(J_1, C_1, J_2, C_2, j_1, c_1, j_2, c_2)^T$, $B=(0, 0, 0, 0, 0, 0, 0, 1)^T$ and
$$
A=\left[\begin{array}{cccccccc}
0& G_1(0)&F_2(L)&-G_2(L)&0&0&0&0\cr
-F_1(a)&G_1(a)&0&-G_2(a)&0&0&0&0\cr
0&0&0&0&0&g_1(0)&f_2(L)&-g_2(L)\cr
0&0&0&0&-f_1(a)&g_1(a)&0&-g_2(a)\cr
1&0&-1&-\omega_aG_2(a)&0&0&0&\omega_dg_2(a)\cr
1&0&-1&0&1&0&-1&0\cr
-IF_1&IG_1&-IF_2&IG_2&-If_1&Ig_1&-If_2&Ig_2
\end{array}\right],
$$
in which $IH_1=\int_0^aH_1dx, IH_2=\int_a^LH_2dx$ for $H=F, G, f, g$.
Although it can be obtained explicitly, the solution of $AX=B$ is very complex. So, for simplicity, we only discuss the special cases in which potentials $V_1, V_2$ satisfy $\Delta V_1=\Delta V_2=0$. By routine analysis, one can obtain
$$
\begin{array}{ll}
J_1=&F_2(L)G_1(a)[-\omega_a\Omega_dG_1(a)g_1(0)f_2(L)g_1(a)-\omega_a\Omega_dG_1(a)f_1(a)g_1(0)^2\cr
&\Omega_a\omega_dG_1(0)g_1(a)^2f_2(L)+\Omega_a\omega_dG_1(0)f_1(a)g_1(0)g_1(a)]/\det(A),
\end{array}
$$
$$
\begin{array}{ll}
J_2=&-F_1(a)G_1(0)[-\omega_a\Omega_dG_1(a)g_1(0)f_2(L)g_1(a)-\omega_a\Omega_dG_1(a)f_1(a)g_1(0)^2\cr
&\Omega_a\omega_dG_1(0)g_1(a)^2f_2(L)+\Omega_a\omega_dG_1(0)f_1(a)g_1(0)g_1(a)]/\det(A),
\end{array}
$$
$$
\begin{array}{ll}
j_1=&-f_2(L)g_1(a)[-\omega_a\Omega_dF_1(a)G_1(0)G_1(a)g_1(0)-\omega_a\Omega_dF_2(L)G_1(a)^2g_1(0)\cr
&+\Omega_a\omega_dF_1(a)G_1(0)^2g_1(a)+\Omega_a\omega_dG_1(0)F_2(L)G_1(a)g_1(a)]/\det(A),
\end{array}
$$
$$
\begin{array}{ll}
j_2=&f_1(a)g_1(0)[-\omega_a\Omega_dF_1(a)G_1(0)G_1(a)g_1(0)-\omega_a\Omega_dF_2(L)G_1(a)^2g_1(0)\cr
&+\Omega_a\omega_dF_1(a)G_1(0)^2g_1(a)+\Omega_a\omega_dG_1(0)F_2(L)G_1(a)g_1(a)]/\det(A),
\end{array}
$$
where $\det(A)$ is the determinant of matrix $A$, and it can be proved that $\det(A)<0$.
So the total probability flux is
\begin{equation}\label{eq71}
\begin{aligned}
J_1+j_1=&J_2+j_2\cr
=&[\Omega_a\omega_dG_1(0)g_1(a)-\omega_a\Omega_dG_1(a)g_1(0)]\cr
&\times[f_1(a)g_1(0)F_2(L)G_1(a)-F_1(a)G_1(0)f_2(L)g_1(a)]/\det(A)\cr
=&\frac{[g_1(0)G_1(0)]^2f_1(a)F_1(a)}{\det(A)}\left[\Omega_a\omega_d\frac{g_1(a)}{g_1(0)}-\omega_a\Omega_d\frac{G_1(a)}{G_1(0)}\right]\cr
&\times\left[\frac{F_2(L)}{F_1(a)}\frac{G_1(a)}{G_1(0)}-\frac{f_2(L)}{f_1(a)}\frac{g_1(a)}{g_1(0)}\right]\cr
=&\frac{[g_1(0)G_1(0)]^2f_1(a)F_1(a)}{\det(A)}\left[\Omega_a\omega_de^{\beta(V_2(0)-V_2(a))}-\omega_a\Omega_de^{\beta(V_1(0)-V_1(a))}\right]\cr
&\times\left[\frac{\int_a^Le^{\beta V_1(y)}}{\int_0^ae^{\beta V_1(y)}}-\frac{\int_a^Le^{\beta V_2(y)}}{\int_0^ae^{\beta V_2(y)}}\right].
\end{aligned}
\end{equation}
Obviously, $J_i+j_i>0$ if and only if $\left[\Omega_a\omega_de^{\beta(V_2(0)-V_2(a))}-\omega_a\Omega_de^{\beta(V_1(0)-V_1(a))}\right]\times$ $\left[\frac{\int_a^Le^{\beta V_1(y)}}{\int_0^ae^{\beta V_1(y)}}-\frac{\int_a^Le^{\beta V_2(y)}}{\int_0^ae^{\beta V_2(y)}}\right]<0$. In view of the expression in (\ref{eq46}), one can find that, the properties of this special case are similar as those of the special case {\bf II} of the coupled one-dimensional hopping models. The mean velocity of molecular motor might not be zero even if there is no energy input in each state. For this special case, the energy for motor motion comes from the processes that drive the motor from one state to another \cite{Parmeggiani1999, Astumian1997, Parrondo2002, Reimann20021}.

\subsection{Special case III: $\omega_a(x)=\omega_d(x)\equiv 0$ for $x\ne a, L$, and $V_2(x)$ is constant}
For this special case, the governing equations of steady state probability densities $P(x), \rho(x)$ are as follows
\begin{equation}\label{eq72}
\begin{aligned}
&\left\{\begin{aligned}
&D\partial_x(\beta
V_1' P_1+\partial_x P_1)=0,\cr
&D\partial^2_x\rho_1=0,
\end{aligned}\right.\quad 0< x< a,\cr
&\left\{\begin{aligned}
&D\partial_x(\beta
V_1' P_2+\partial_x P_2)=0,\cr
&D\partial^2_x\rho_2=0,
\end{aligned}\right.\quad a< x< L.
\end{aligned}
\end{equation}
Its general solutions are (\ref{eq68}) but with $f_i(x)=x/D, g_i(x)\equiv 1$. The solution which satisfies the constraints (\ref{eq67}) is as follows
$$
\begin{aligned}
&\begin{array}{ll}
J_1=&-2[\omega_dG_1(0)G_2(a)DL-\omega_dG_1(a)G_2(L)DL+\Omega_dG_1(0)G_2(a)DL\cr
 &-\Omega_dG_1(a)G_2(L)DL+\Omega_d\omega_dG_1(0)G_2(a)aL-\Omega_d\omega_dG_1(0)G_2(a)a^2\cr
&-\Omega_d\omega_dG_1(a)G_2(L)aL+\Omega_d\omega_dG_1(a)G_2(L)a^2\cr
&-\omega_a\Omega_dG_1(a)F_2(L)G_2(a)DL+\Omega_a\omega_dG_1(0)F_2(L)G_2(a)LD]/\det(A),
\end{array}\cr
&\begin{array}{ll}
J_2=&-2[\omega_a\Omega_dF_1(a)G_1(0)LDG_2(a)-\Omega_a\omega_dF_1(a)G_1(0)LDG_2(L)\cr
&+\omega_dG_1(0)G_2(a)DL-\omega_dG_1(a)G_2(L)DL+\Omega_dG_1(0)G_2(a)DL\cr
&-\Omega_dG_1(a)G_2(L)DL+\Omega_d\omega_dG_1(0)G_2(a)aL\cr
&-\Omega_d\omega_dG_1(a)G_2(L)aL-\Omega_d\omega_dG_1(0)G_2(a)a^2+\Omega_d\omega_dG_1(a)G_2(L)a^2]/\det(A),
\end{array}\cr
&\begin{array}{ll}
j_1=&2(L-a)[-\omega_a\Omega_dF_1(a)G_1(0)G_2(a)-\omega_a\Omega_dG_1(a)F_2(L)G_2(a)\cr
&+\Omega_a\omega_dF_1(a)G_1(0)G_2(L)+\Omega_a\omega_dG_1(0)F_2(L)G_2(a)]D/\det(A),
\end{array}\cr
&\begin{array}{ll}
j_2=&-2a[-\omega_a\Omega_dF_1(a)G_1(0)G_2(a)-\omega_a\Omega_dG_1(a)F_2(L)G_2(a)\cr
&+\Omega_a\omega_dF_1(a)G_1(0)G_2(L)+\Omega_a\omega_dG_1(0)F_2(L)G_2(a)]D/\det(A).
\end{array}
\end{aligned}
$$
So the total probability flux is
\begin{equation}\label{eq73}
\begin{aligned}
J_1+j_1=&J_2+j_2\cr
=&2\{[G_1(a)G_2(L)-G_1(0)G_2(a)][\omega_dDL+\Omega_dDL+\omega_d\Omega_d(L-a)a]\cr
&+2aDF_2(L)G_2(a)[\omega_a\Omega_dG_1(a)-\Omega_a\omega_dG_1(0)]\cr
&+2D(L-a)F_1(a)G_1(0)[\Omega_a\omega_dG_2(L)-\omega_a\Omega_dG_2(a)]\}/\det(A).
\end{aligned}
\end{equation}
More specially, if potential $V_1(x)$ is periodic and continuous at $a$, then $G_1(0)=G_2(L), G_1(a)=G_2(a)$. So
\begin{equation}\label{eq74}
\begin{aligned}
J_1+j_1=&\frac{2D}{\det(A)}[\omega_a\Omega_dG_1(a)-\Omega_a\omega_dG_1(0)][aF_2(L)G_2(a)-(L-a)F_1(a)G_1(0)]\cr
=&\frac{2D(G_1(0))^2G_2(a)}{\det(A)}[\omega_a\Omega_de^{\beta (V_1(0)-V_1(a))}-\Omega_a\omega_d]\cr
&\times\left(a\int_a^Le^{\beta V_1(x)}dx-(L-a)\int_0^ae^{\beta V_1(x)}dx\right)\cr
=&\frac{2D(G_1(0))^2G_2(a)}{\det(A)}[\omega_a\Omega_de^{\beta (V_1(0)-V_1(a))}-\Omega_a\omega_d]\cr
&\times\left(a\int_0^Le^{\beta V_1(x)}dx-L\int_0^ae^{\beta V_1(x)}dx\right)\cr
=&\frac{2LD(G_1(0))^2G_2(a)\int_0^Le^{\beta V_1(x)}dx}{\det(A)}[\Omega_a\omega_d-\omega_a\Omega_de^{\beta (V_1(0)-V_1(a))}]\cr
&\times\left(\frac{L-a}{L}-\frac{\int_a^Le^{\beta V_1(x)}dx}{\int_0^Le^{\beta V_1(x)}dx}\right)
\end{aligned}
\end{equation}
Therefore, the total probability flux $J_1+j_1>0$ if and only if $[\omega_a\Omega_de^{\beta (V_1(0)-V_1(a))}-\Omega_a\omega_d]\left(a\int_0^Le^{\beta V_1(x)}dx-L\int_0^ae^{\beta V_1(x)}dx\right)<0$. Similar as before, from (\ref{eq52}) and (\ref{eq74}) one can find the similarity between them \cite{Zhang2010}.

\begin{figure}
  \includegraphics[width=130pt]{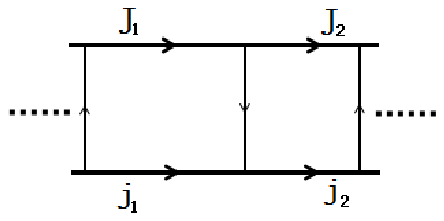}\includegraphics[width=130pt]{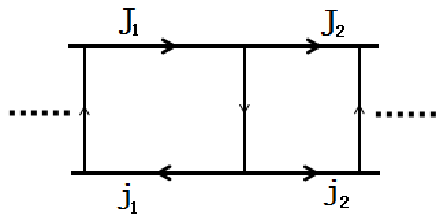}\includegraphics[width=130pt]{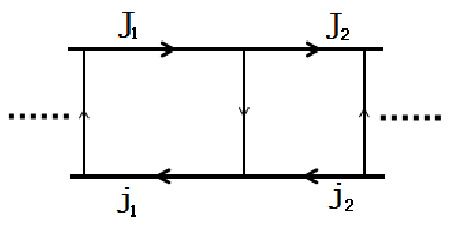}\\
  \includegraphics[width=130pt]{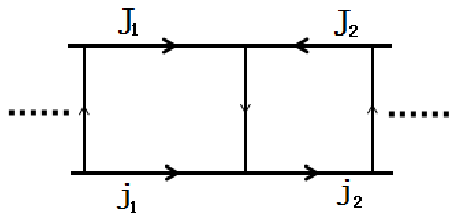}\includegraphics[width=130pt]{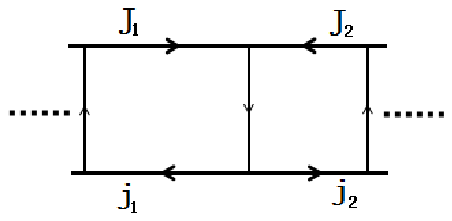}\includegraphics[width=130pt]{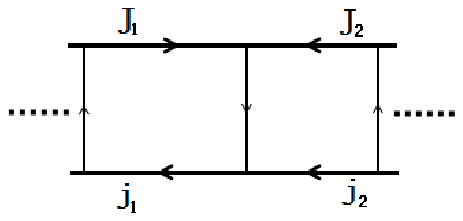}\\
  \includegraphics[width=130pt]{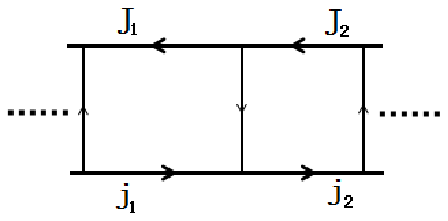}\includegraphics[width=130pt]{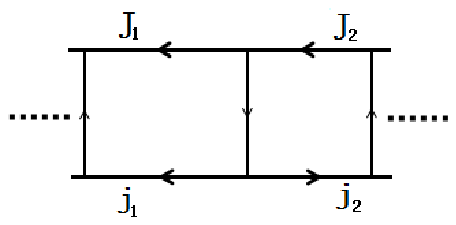}\includegraphics[width=130pt]{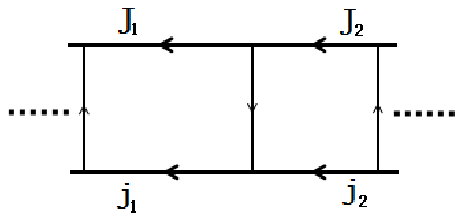}\\
  \caption{Different types of probability flux for special case {\bf II}. From these figures, one can see that the motion of molecular motors is usually loosely coupled to the energy input process, i.e., there might exist energy input but without directed macroscopic mechanical motion. Part of the input energy will be consumed during substep oscillation.}\label{Fig5}
\end{figure}

To better understand the properties of our model, we discuss the direction of probability fluxes here.
For the special case in which there are only two locations at which the inter-state transition rates are nonzero, i.e., the special case {\bf II}, there are altogether $18$ different types of probability flux. Since the states 1 and 2 are temporally symmetric, we restrict our discussion only on the cases in which $\omega_aP(a)-\omega_d\rho(a)\ge 0$ for the continuous model, or $\omega_aP_M-\omega_d\rho_M\ge 0$ for the hopping model. Then, there are altogether $9$ different types of probability flux (see Fig. \ref{Fig5}). Furthermore, if $\Delta V_1=\Delta V_2=0$, then there is only one type (see the figure (2, 2) in Fig. \ref{Fig5}).
On the other hand, if the potential $V_2$ is constant (or $f_i=b_i\equiv f$ for the hopping model), then there are altogether 3 different types of probability flux (see the second column in Fig. \ref{Fig5}).

\section{Conclusions}
In conclusion, two chemical states models of molecular motor are discussed in this paper. For some special cases, explicit expressions of mean velocity are obtained. We find that the mean velocity of molecular motor might not be zero even if both of the potentials are periodic, which means there is no energy input to the molecular motor in each of the chemical states. The energy for the motion molecular motor motion comes from the processes that drive the motor from one state to another. For motor proteins, these processes are ATP hydrolysis. At the same time, from the expression of mean velocity, we find that the velocity of molecular motor might be zero even if there exists nonzero input energy. Which implies that the motion of motor protein is usually loosely coupled to ATP hydrolysis \cite{Bieling2008, Endres2006, Seidel2008, Shaevitz2005, Yildiz2008, Gao2006, Nishikawa2008, Masuda2009, Gerritsma2009}.

\vskip 0.5cm

\acknowledgments{The author has been funded by the National Natural Science Foundation of China (under Grant No. 10701029). He is grateful to the China Scholarship Council for their financial support of his study in United States and thanks Professor Michael E Fisher, Devarajan Thirumalai, and the Institute for Physical Science and Technology at the University of Maryland for their hospitality.}

\newpage



\begin{thebibliography}{107}
\expandafter\ifx\csname natexlab\endcsname\relax\def\natexlab#1{#1}\fi
\expandafter\ifx\csname bibnamefont\endcsname\relax
  \def\bibnamefont#1{#1}\fi
\expandafter\ifx\csname bibfnamefont\endcsname\relax
  \def\bibfnamefont#1{#1}\fi
\expandafter\ifx\csname citenamefont\endcsname\relax
  \def\citenamefont#1{#1}\fi
\expandafter\ifx\csname url\endcsname\relax
  \def\url#1{\texttt{#1}}\fi
\expandafter\ifx\csname urlprefix\endcsname\relax\def\urlprefix{URL }\fi
\providecommand{\bibinfo}[2]{#2}
\providecommand{\eprint}[2][]{\url{#2}}

\bibitem[{\citenamefont{Bray}(2001)}]{Bray2001}
\bibinfo{author}{\bibfnamefont{D.}~\bibnamefont{Bray}},
  \emph{\bibinfo{title}{Cell movements: from molecules to motility, 2nd Edn}}
  (\bibinfo{publisher}{Garland, New York}, \bibinfo{year}{2001}).

\bibitem[{\citenamefont{Howard}(2001)}]{Howard2001}
\bibinfo{author}{\bibfnamefont{J.}~\bibnamefont{Howard}},
  \emph{\bibinfo{title}{Mechanics of Motor Proteins and the Cytoskeleton}}
  (\bibinfo{publisher}{Sinauer Associates and Sunderland, MA},
  \bibinfo{year}{2001}).

\bibitem[{\citenamefont{Badoual et~al.}(2002)\citenamefont{Badoual,
  J\"{u}licher, and Prost}}]{Badoual2002}
\bibinfo{author}{\bibfnamefont{M.}~\bibnamefont{Badoual}},
  \bibinfo{author}{\bibfnamefont{F.}~\bibnamefont{J\"{u}licher}},
  \bibnamefont{and} \bibinfo{author}{\bibfnamefont{J.}~\bibnamefont{Prost}},
  \bibinfo{journal}{Proc. Natl. Acad. Sci. USA} \textbf{\bibinfo{volume}{99}},
  \bibinfo{pages}{6696} (\bibinfo{year}{2002}).

\bibitem[{\citenamefont{Klumpp and Lipowsky}(2005)}]{Lipowsky2005}
\bibinfo{author}{\bibfnamefont{S.}~\bibnamefont{Klumpp}} \bibnamefont{and}
  \bibinfo{author}{\bibfnamefont{R.}~\bibnamefont{Lipowsky}},
  \bibinfo{journal}{Proc. Natl. Acad. Sci. USA} \textbf{\bibinfo{volume}{102}},
  \bibinfo{pages}{17284} (\bibinfo{year}{2005}).

\bibitem[{\citenamefont{Riedel-Kruse et~al.}(2007)\citenamefont{Riedel-Kruse,
  Hilfinger, Howard, and J\"{u}licher}}]{Riedel2007}
\bibinfo{author}{\bibfnamefont{I.~H.} \bibnamefont{Riedel-Kruse}},
  \bibinfo{author}{\bibfnamefont{A.}~\bibnamefont{Hilfinger}},
  \bibinfo{author}{\bibfnamefont{J.}~\bibnamefont{Howard}}, \bibnamefont{and}
  \bibinfo{author}{\bibfnamefont{F.}~\bibnamefont{J\"{u}licher}},
  \bibinfo{journal}{HFSP J.} \textbf{\bibinfo{volume}{1}}, \bibinfo{pages}{192}
  (\bibinfo{year}{2007}).

\bibitem[{\citenamefont{Zhang}(2009{\natexlab{a}})}]{Zhang2009}
\bibinfo{author}{\bibfnamefont{Y.}~\bibnamefont{Zhang}},
  \bibinfo{journal}{Phys. Rev. E} \textbf{\bibinfo{volume}{79}},
  \bibinfo{pages}{061918} (\bibinfo{year}{2009}{\natexlab{a}}).

\bibitem[{\citenamefont{Howard}(2009)}]{Howard2009}
\bibinfo{author}{\bibfnamefont{J.}~\bibnamefont{Howard}},
  \bibinfo{journal}{Annu. Rev. Biophys.} \textbf{\bibinfo{volume}{38}},
  \bibinfo{pages}{217} (\bibinfo{year}{2009}).

\bibitem[{\citenamefont{Vale}(2003)}]{Vale2003}
\bibinfo{author}{\bibfnamefont{R.~D.} \bibnamefont{Vale}},
  \bibinfo{journal}{Cell} \textbf{\bibinfo{volume}{112}}, \bibinfo{pages}{467}
  (\bibinfo{year}{2003}).

\bibitem[{\citenamefont{Fisher and Kolomeisky}(2001)}]{Fisher2001}
\bibinfo{author}{\bibfnamefont{M.~E.} \bibnamefont{Fisher}} \bibnamefont{and}
  \bibinfo{author}{\bibfnamefont{A.~B.} \bibnamefont{Kolomeisky}},
  \bibinfo{journal}{Proc. Natl. Acad. Sci. USA} \textbf{\bibinfo{volume}{98}},
  \bibinfo{pages}{7748} (\bibinfo{year}{2001}).

\bibitem[{\citenamefont{Carter and Cross}(2005)}]{Carter2005}
\bibinfo{author}{\bibfnamefont{N.~J.} \bibnamefont{Carter}} \bibnamefont{and}
  \bibinfo{author}{\bibfnamefont{R.~A.} \bibnamefont{Cross}},
  \bibinfo{journal}{Nature} \textbf{\bibinfo{volume}{435}},
  \bibinfo{pages}{308} (\bibinfo{year}{2005}).

\bibitem[{\citenamefont{Block}(2007)}]{Block2007}
\bibinfo{author}{\bibfnamefont{S.~M.} \bibnamefont{Block}},
  \bibinfo{journal}{Biophys. J.} \textbf{\bibinfo{volume}{92}},
  \bibinfo{pages}{2986} (\bibinfo{year}{2007}).

\bibitem[{\citenamefont{Zhang}(2008)}]{Zhang2008}
\bibinfo{author}{\bibfnamefont{Y.}~\bibnamefont{Zhang}},
  \bibinfo{journal}{Biophys. Chem.} \textbf{\bibinfo{volume}{136}},
  \bibinfo{pages}{19} (\bibinfo{year}{2008}).

\bibitem[{\citenamefont{Toprak et~al.}(2009)\citenamefont{Toprak, Yildiz,
  Hoffman, Rosenfeld, and Selvin}}]{Toprak2009}
\bibinfo{author}{\bibfnamefont{E.}~\bibnamefont{Toprak}},
  \bibinfo{author}{\bibfnamefont{A.}~\bibnamefont{Yildiz}},
  \bibinfo{author}{\bibfnamefont{M.~T.} \bibnamefont{Hoffman}},
  \bibinfo{author}{\bibfnamefont{S.~S.} \bibnamefont{Rosenfeld}},
  \bibnamefont{and} \bibinfo{author}{\bibfnamefont{P.~R.}
  \bibnamefont{Selvin}}, \bibinfo{journal}{Proc. Natl. Acad. Sci. USA}
  \textbf{\bibinfo{volume}{106}}, \bibinfo{pages}{12717}
  (\bibinfo{year}{2009}).

\bibitem[{\citenamefont{Guydosh and Block}(2009)}]{Guydosh2009}
\bibinfo{author}{\bibfnamefont{N.~R.} \bibnamefont{Guydosh}} \bibnamefont{and}
  \bibinfo{author}{\bibfnamefont{S.~M.} \bibnamefont{Block}},
  \bibinfo{journal}{Nature} \textbf{\bibinfo{volume}{08259}}
  (\bibinfo{year}{2009}).

\bibitem[{\citenamefont{Hyeon et~al.}(2009)\citenamefont{Hyeon, Klumppb, and
  Onuchic}}]{Hyeon2009}
\bibinfo{author}{\bibfnamefont{C.}~\bibnamefont{Hyeon}},
  \bibinfo{author}{\bibfnamefont{S.}~\bibnamefont{Klumppb}}, \bibnamefont{and}
  \bibinfo{author}{\bibfnamefont{J.~N.} \bibnamefont{Onuchic}},
  \bibinfo{journal}{Phys. Chem. Chem. Phys} \textbf{\bibinfo{volume}{11}},
  \bibinfo{pages}{4899} (\bibinfo{year}{2009}).

\bibitem[{\citenamefont{Hariharan and Hancock}(2009)}]{Hariharan2009}
\bibinfo{author}{\bibfnamefont{V.}~\bibnamefont{Hariharan}} \bibnamefont{and}
  \bibinfo{author}{\bibfnamefont{W.~O.} \bibnamefont{Hancock}},
  \bibinfo{journal}{Cell. Mol. Bioe.} \textbf{\bibinfo{volume}{2}},
  \bibinfo{pages}{177} (\bibinfo{year}{2009}).

\bibitem[{\citenamefont{Reck-Peterson et~al.}(2006)\citenamefont{Reck-Peterson,
  Yildiz, Carter, Gennerich, Zhang, and Vale}}]{Samara2006}
\bibinfo{author}{\bibfnamefont{S.~L.} \bibnamefont{Reck-Peterson}},
  \bibinfo{author}{\bibfnamefont{A.}~\bibnamefont{Yildiz}},
  \bibinfo{author}{\bibfnamefont{A.~P.} \bibnamefont{Carter}},
  \bibinfo{author}{\bibfnamefont{A.}~\bibnamefont{Gennerich}},
  \bibinfo{author}{\bibfnamefont{N.}~\bibnamefont{Zhang}}, \bibnamefont{and}
  \bibinfo{author}{\bibfnamefont{R.~D.} \bibnamefont{Vale}},
  \bibinfo{journal}{Cell} \textbf{\bibinfo{volume}{126}}, \bibinfo{pages}{335}
  (\bibinfo{year}{2006}).

\bibitem[{\citenamefont{Toba et~al.}(2006)\citenamefont{Toba, Watanabe,
  Yamaguchi-Okimoto, Toyoshima, and Higuchi}}]{Toba2006}
\bibinfo{author}{\bibfnamefont{S.}~\bibnamefont{Toba}},
  \bibinfo{author}{\bibfnamefont{T.~M.} \bibnamefont{Watanabe}},
  \bibinfo{author}{\bibfnamefont{L.}~\bibnamefont{Yamaguchi-Okimoto}},
  \bibinfo{author}{\bibfnamefont{Y.~Y.} \bibnamefont{Toyoshima}},
  \bibnamefont{and} \bibinfo{author}{\bibfnamefont{H.}~\bibnamefont{Higuchi}},
  \bibinfo{journal}{Proc. Natl. Acad. Sci. USA} \textbf{\bibinfo{volume}{103}},
  \bibinfo{pages}{5741} (\bibinfo{year}{2006}).

\bibitem[{\citenamefont{Gennerich and Vale}(2009)}]{Gennerich2009}
\bibinfo{author}{\bibfnamefont{A.}~\bibnamefont{Gennerich}} \bibnamefont{and}
  \bibinfo{author}{\bibfnamefont{R.~D.} \bibnamefont{Vale}},
  \bibinfo{journal}{Curr. Opin. Cell. Biol.} \textbf{\bibinfo{volume}{21}},
  \bibinfo{pages}{59} (\bibinfo{year}{2009}).

\bibitem[{\citenamefont{Houdusse and Carter}(2009)}]{Houdusse2009}
\bibinfo{author}{\bibfnamefont{A.}~\bibnamefont{Houdusse}} \bibnamefont{and}
  \bibinfo{author}{\bibfnamefont{A.~P.} \bibnamefont{Carter}},
  \bibinfo{journal}{Cell} \textbf{\bibinfo{volume}{136}}, \bibinfo{pages}{395}
  (\bibinfo{year}{2009}).

\bibitem[{\citenamefont{Roberts et~al.}(2009)\citenamefont{Roberts, Numata,
  Walker, Kato, Malkova, Kon, Ohkura, Arisaka, Knight, Sutoh
  et~al.}}]{Roberts2009}
\bibinfo{author}{\bibfnamefont{A.~J.} \bibnamefont{Roberts}},
  \bibinfo{author}{\bibfnamefont{N.}~\bibnamefont{Numata}},
  \bibinfo{author}{\bibfnamefont{M.~L.} \bibnamefont{Walker}},
  \bibinfo{author}{\bibfnamefont{Y.~S.} \bibnamefont{Kato}},
  \bibinfo{author}{\bibfnamefont{B.}~\bibnamefont{Malkova}},
  \bibinfo{author}{\bibfnamefont{T.}~\bibnamefont{Kon}},
  \bibinfo{author}{\bibfnamefont{R.}~\bibnamefont{Ohkura}},
  \bibinfo{author}{\bibfnamefont{F.}~\bibnamefont{Arisaka}},
  \bibinfo{author}{\bibfnamefont{P.~J.} \bibnamefont{Knight}},
  \bibinfo{author}{\bibfnamefont{K.}~\bibnamefont{Sutoh}},
  \bibnamefont{et~al.}, \bibinfo{journal}{Cell} \textbf{\bibinfo{volume}{136}},
  \bibinfo{pages}{485} (\bibinfo{year}{2009}).

\bibitem[{\citenamefont{Kardon et~al.}(2009)\citenamefont{Kardon,
  Reck-Peterson, and Vale}}]{Kardona2009}
\bibinfo{author}{\bibfnamefont{J.~R.} \bibnamefont{Kardon}},
  \bibinfo{author}{\bibfnamefont{S.~L.} \bibnamefont{Reck-Peterson}},
  \bibnamefont{and} \bibinfo{author}{\bibfnamefont{R.~D.} \bibnamefont{Vale}},
  \bibinfo{journal}{Proc. Natl. Acad. Sci. USA} \textbf{\bibinfo{volume}{106}},
  \bibinfo{pages}{5669} (\bibinfo{year}{2009}).

\bibitem[{\citenamefont{Serohijos et~al.}(2009)\citenamefont{Serohijos,
  Tsygankov, Liu, Elstonbd, and Dokholyanz}}]{Serohijos2009}
\bibinfo{author}{\bibfnamefont{A.~W.~R.} \bibnamefont{Serohijos}},
  \bibinfo{author}{\bibfnamefont{W.~D.} \bibnamefont{Tsygankov}},
  \bibinfo{author}{\bibfnamefont{S.}~\bibnamefont{Liu}},
  \bibinfo{author}{\bibfnamefont{T.~C.} \bibnamefont{Elstonbd}},
  \bibnamefont{and} \bibinfo{author}{\bibfnamefont{N.~V.}
  \bibnamefont{Dokholyanz}}, \bibinfo{journal}{Phys. Chem. Chem. Phys.}
  \textbf{\bibinfo{volume}{11}}, \bibinfo{pages}{4840} (\bibinfo{year}{2009}).

\bibitem[{\citenamefont{Rosenfeld and Sweeney}(2004)}]{Rosenfeld2004}
\bibinfo{author}{\bibfnamefont{S.~S.} \bibnamefont{Rosenfeld}}
  \bibnamefont{and} \bibinfo{author}{\bibfnamefont{H.~L.}
  \bibnamefont{Sweeney}}, \bibinfo{journal}{J. Biol. Chem.}
  \textbf{\bibinfo{volume}{279}}, \bibinfo{pages}{40100}
  (\bibinfo{year}{2004}).

\bibitem[{\citenamefont{Purcell et~al.}(2005)\citenamefont{Purcell, Sweeney,
  and Spudich}}]{Purcell2005}
\bibinfo{author}{\bibfnamefont{T.~J.} \bibnamefont{Purcell}},
  \bibinfo{author}{\bibfnamefont{H.~L.} \bibnamefont{Sweeney}},
  \bibnamefont{and} \bibinfo{author}{\bibfnamefont{J.~A.}
  \bibnamefont{Spudich}}, \bibinfo{journal}{Proc. Natl. Acad. Sci. USA}
  \textbf{\bibinfo{volume}{102}}, \bibinfo{pages}{13873}
  (\bibinfo{year}{2005}).

\bibitem[{\citenamefont{Veigel et~al.}(2003)\citenamefont{Veigel, Molloy,
  Schmitz, and Kendrick-Jones}}]{Veigel2005}
\bibinfo{author}{\bibfnamefont{C.}~\bibnamefont{Veigel}},
  \bibinfo{author}{\bibfnamefont{J.~E.} \bibnamefont{Molloy}},
  \bibinfo{author}{\bibfnamefont{S.}~\bibnamefont{Schmitz}}, \bibnamefont{and}
  \bibinfo{author}{\bibfnamefont{J.}~\bibnamefont{Kendrick-Jones}},
  \bibinfo{journal}{Nat. Cell. Biol.} \textbf{\bibinfo{volume}{5}},
  \bibinfo{pages}{980} (\bibinfo{year}{2003}).

\bibitem[{\citenamefont{Sakamoto et~al.}(2008)\citenamefont{Sakamoto, Webb,
  Forgacs, White, and Sellers}}]{Sakamoto2008}
\bibinfo{author}{\bibfnamefont{T.}~\bibnamefont{Sakamoto}},
  \bibinfo{author}{\bibfnamefont{M.~R.} \bibnamefont{Webb}},
  \bibinfo{author}{\bibfnamefont{E.}~\bibnamefont{Forgacs}},
  \bibinfo{author}{\bibfnamefont{H.~D.} \bibnamefont{White}}, \bibnamefont{and}
  \bibinfo{author}{\bibfnamefont{J.~R.} \bibnamefont{Sellers}},
  \bibinfo{journal}{Nature} \textbf{\bibinfo{volume}{455}},
  \bibinfo{pages}{128} (\bibinfo{year}{2008}).

\bibitem[{\citenamefont{Del R.~Jackson and Baker}(2009)}]{Jackson2009}
\bibinfo{author}{\bibfnamefont{J.}~\bibnamefont{Del R.~Jackson}}
  \bibnamefont{and} \bibinfo{author}{\bibfnamefont{J.~E.} \bibnamefont{Baker}},
  \bibinfo{journal}{Phys. Chem. Chem. Phys.} \textbf{\bibinfo{volume}{11}},
  \bibinfo{pages}{4808} (\bibinfo{year}{2009}).

\bibitem[{\citenamefont{Fedorov et~al.}(2009)\citenamefont{Fedorov, B\"{o}hl,
  Tsiavaliaris, Hartmann, Taft, Baruch, Brenner, Martin, Kn\"{o}lker, Gutzeit
  et~al.}}]{Fedorov2009}
\bibinfo{author}{\bibfnamefont{R.}~\bibnamefont{Fedorov}},
  \bibinfo{author}{\bibfnamefont{M.}~\bibnamefont{B\"{o}hl}},
  \bibinfo{author}{\bibfnamefont{G.}~\bibnamefont{Tsiavaliaris}},
  \bibinfo{author}{\bibfnamefont{F.~K.} \bibnamefont{Hartmann}},
  \bibinfo{author}{\bibfnamefont{M.~H.} \bibnamefont{Taft}},
  \bibinfo{author}{\bibfnamefont{P.}~\bibnamefont{Baruch}},
  \bibinfo{author}{\bibfnamefont{B.}~\bibnamefont{Brenner}},
  \bibinfo{author}{\bibfnamefont{R.}~\bibnamefont{Martin}},
  \bibinfo{author}{\bibfnamefont{H.-J.} \bibnamefont{Kn\"{o}lker}},
  \bibinfo{author}{\bibfnamefont{H.~O.} \bibnamefont{Gutzeit}},
  \bibnamefont{et~al.}, \bibinfo{journal}{Nat. Struct. Mol. Biol.}
  \textbf{\bibinfo{volume}{16}}, \bibinfo{pages}{80} (\bibinfo{year}{2009}).

\bibitem[{\citenamefont{Wang and Oster}(1998)}]{Wang1998}
\bibinfo{author}{\bibfnamefont{H.}~\bibnamefont{Wang}} \bibnamefont{and}
  \bibinfo{author}{\bibfnamefont{G.}~\bibnamefont{Oster}},
  \bibinfo{journal}{Nature} \textbf{\bibinfo{volume}{396}},
  \bibinfo{pages}{279} (\bibinfo{year}{1998}).

\bibitem[{\citenamefont{Kinosita et~al.}(2000)\citenamefont{Kinosita, H.Noji,
  K.Adachi, and Yasuda}}]{Kazuhiko2000}
\bibinfo{author}{\bibfnamefont{K.}~\bibnamefont{Kinosita}},
  \bibinfo{author}{\bibnamefont{H.Noji}},
  \bibinfo{author}{\bibnamefont{K.Adachi}}, \bibnamefont{and}
  \bibinfo{author}{\bibfnamefont{R.}~\bibnamefont{Yasuda}},
  \bibinfo{journal}{Phil. Trans. R. Soc. B} \textbf{\bibinfo{volume}{355}},
  \bibinfo{pages}{473} (\bibinfo{year}{2000}).

\bibitem[{\citenamefont{Nishizaka et~al.}(2004)\citenamefont{Nishizaka, Oiwa,
  Noji, Kimura, Muneyuki, Yoshida, and Jr}}]{Nishizaka2004}
\bibinfo{author}{\bibfnamefont{T.}~\bibnamefont{Nishizaka}},
  \bibinfo{author}{\bibfnamefont{K.}~\bibnamefont{Oiwa}},
  \bibinfo{author}{\bibfnamefont{H.}~\bibnamefont{Noji}},
  \bibinfo{author}{\bibfnamefont{S.}~\bibnamefont{Kimura}},
  \bibinfo{author}{\bibfnamefont{E.}~\bibnamefont{Muneyuki}},
  \bibinfo{author}{\bibfnamefont{M.}~\bibnamefont{Yoshida}}, \bibnamefont{and}
  \bibinfo{author}{\bibfnamefont{K.~K.} \bibnamefont{Jr}},
  \bibinfo{journal}{Nat. Struct. Mol. Biol.} \textbf{\bibinfo{volume}{11}},
  \bibinfo{pages}{142} (\bibinfo{year}{2004}).

\bibitem[{\citenamefont{Adachi et~al.}(2007)\citenamefont{Adachi, Oiwa,
  Nishizaka, Furuike, Noji, Itoh, Yoshida, and K.~Kinosita}}]{Adachi2007}
\bibinfo{author}{\bibfnamefont{K.}~\bibnamefont{Adachi}},
  \bibinfo{author}{\bibfnamefont{K.}~\bibnamefont{Oiwa}},
  \bibinfo{author}{\bibfnamefont{T.}~\bibnamefont{Nishizaka}},
  \bibinfo{author}{\bibfnamefont{S.}~\bibnamefont{Furuike}},
  \bibinfo{author}{\bibfnamefont{H.}~\bibnamefont{Noji}},
  \bibinfo{author}{\bibfnamefont{H.}~\bibnamefont{Itoh}},
  \bibinfo{author}{\bibfnamefont{M.}~\bibnamefont{Yoshida}}, \bibnamefont{and}
  \bibinfo{author}{\bibfnamefont{J.}~\bibnamefont{K.~Kinosita}},
  \bibinfo{journal}{Cell} \textbf{\bibinfo{volume}{130}}, \bibinfo{pages}{309}
  (\bibinfo{year}{2007}).

\bibitem[{\citenamefont{Muneyuki et~al.}(2007)\citenamefont{Muneyuki,
  Watanabe-Nakayama, Suzuki, Yoshida, Nishizaka, and Noji}}]{Muneyuki2007}
\bibinfo{author}{\bibfnamefont{E.}~\bibnamefont{Muneyuki}},
  \bibinfo{author}{\bibfnamefont{T.}~\bibnamefont{Watanabe-Nakayama}},
  \bibinfo{author}{\bibfnamefont{T.}~\bibnamefont{Suzuki}},
  \bibinfo{author}{\bibfnamefont{M.}~\bibnamefont{Yoshida}},
  \bibinfo{author}{\bibfnamefont{T.}~\bibnamefont{Nishizaka}},
  \bibnamefont{and} \bibinfo{author}{\bibfnamefont{H.}~\bibnamefont{Noji}},
  \bibinfo{journal}{Biophys. J.} \textbf{\bibinfo{volume}{92}},
  \bibinfo{pages}{1806} (\bibinfo{year}{2007}).

\bibitem[{\citenamefont{Junge et~al.}(2009)\citenamefont{Junge, Sielaff, and
  Engelbrecht}}]{Junge2009}
\bibinfo{author}{\bibfnamefont{W.}~\bibnamefont{Junge}},
  \bibinfo{author}{\bibfnamefont{H.}~\bibnamefont{Sielaff}}, \bibnamefont{and}
  \bibinfo{author}{\bibfnamefont{S.}~\bibnamefont{Engelbrecht}},
  \bibinfo{journal}{Nature} \textbf{\bibinfo{volume}{459}},
  \bibinfo{pages}{364} (\bibinfo{year}{2009}).

\bibitem[{\citenamefont{Jr. et~al.}(2009)\citenamefont{Jr., Vajrala, Infante,
  R.Claycomb, Palanisami, Fang, and Mercier}}]{Miller2009}
\bibinfo{author}{\bibfnamefont{J.~H.~M.} \bibnamefont{Jr.}},
  \bibinfo{author}{\bibfnamefont{V.}~\bibnamefont{Vajrala}},
  \bibinfo{author}{\bibfnamefont{H.~L.} \bibnamefont{Infante}},
  \bibinfo{author}{\bibfnamefont{J.}~\bibnamefont{R.Claycomb}},
  \bibinfo{author}{\bibfnamefont{A.}~\bibnamefont{Palanisami}},
  \bibinfo{author}{\bibfnamefont{J.}~\bibnamefont{Fang}}, \bibnamefont{and}
  \bibinfo{author}{\bibfnamefont{G.~T.} \bibnamefont{Mercier}},
  \bibinfo{journal}{Physica B} \textbf{\bibinfo{volume}{404}},
  \bibinfo{pages}{503} (\bibinfo{year}{2009}).

\bibitem[{\citenamefont{Block et~al.}(1990)\citenamefont{Block, Goldstein, and
  Schnapp}}]{Block1990}
\bibinfo{author}{\bibfnamefont{S.~M.} \bibnamefont{Block}},
  \bibinfo{author}{\bibfnamefont{L.~S.~B.} \bibnamefont{Goldstein}},
  \bibnamefont{and} \bibinfo{author}{\bibfnamefont{B.~J.}
  \bibnamefont{Schnapp}}, \bibinfo{journal}{Nature}
  \textbf{\bibinfo{volume}{348}}, \bibinfo{pages}{348} (\bibinfo{year}{1990}).

\bibitem[{\citenamefont{Yildiz et~al.}(2004)\citenamefont{Yildiz, Tomishige,
  Vale, and Selvin}}]{Yildiz2004}
\bibinfo{author}{\bibfnamefont{A.}~\bibnamefont{Yildiz}},
  \bibinfo{author}{\bibfnamefont{M.}~\bibnamefont{Tomishige}},
  \bibinfo{author}{\bibfnamefont{R.~D.} \bibnamefont{Vale}}, \bibnamefont{and}
  \bibinfo{author}{\bibfnamefont{P.~R.} \bibnamefont{Selvin}},
  \bibinfo{journal}{Science} \textbf{\bibinfo{volume}{303}},
  \bibinfo{pages}{676} (\bibinfo{year}{2004}).

\bibitem[{\citenamefont{Asbury et~al.}(2003)\citenamefont{Asbury, Fehr, and
  Block}}]{Asbury2003}
\bibinfo{author}{\bibfnamefont{C.~L.} \bibnamefont{Asbury}},
  \bibinfo{author}{\bibfnamefont{A.~N.} \bibnamefont{Fehr}}, \bibnamefont{and}
  \bibinfo{author}{\bibfnamefont{S.~M.} \bibnamefont{Block}},
  \bibinfo{journal}{Science} \textbf{\bibinfo{volume}{302}},
  \bibinfo{pages}{2130} (\bibinfo{year}{2003}).

\bibitem[{\citenamefont{Schnitzer and Block}(1997)}]{Schnitzer1997}
\bibinfo{author}{\bibfnamefont{M.~J.} \bibnamefont{Schnitzer}}
  \bibnamefont{and} \bibinfo{author}{\bibfnamefont{S.~M.} \bibnamefont{Block}},
  \bibinfo{journal}{Nature} \textbf{\bibinfo{volume}{388}},
  \bibinfo{pages}{386} (\bibinfo{year}{1997}).

\bibitem[{\citenamefont{Coy et~al.}(1999)\citenamefont{Coy, Wagenbach, and
  Howard}}]{Coy1999}
\bibinfo{author}{\bibfnamefont{D.~L.} \bibnamefont{Coy}},
  \bibinfo{author}{\bibfnamefont{M.}~\bibnamefont{Wagenbach}},
  \bibnamefont{and} \bibinfo{author}{\bibfnamefont{J.}~\bibnamefont{Howard}},
  \bibinfo{journal}{J. Biol. Chem.} \textbf{\bibinfo{volume}{274}},
  \bibinfo{pages}{3667} (\bibinfo{year}{1999}).

\bibitem[{\citenamefont{Fehr et~al.}(2007)\citenamefont{Fehr, Asbury, and
  Block}}]{Fehr2007}
\bibinfo{author}{\bibfnamefont{A.~N.} \bibnamefont{Fehr}},
  \bibinfo{author}{\bibfnamefont{C.~L.} \bibnamefont{Asbury}},
  \bibnamefont{and} \bibinfo{author}{\bibfnamefont{S.~M.} \bibnamefont{Block}},
  \bibinfo{journal}{Biophys. J.}  (\bibinfo{year}{2007}).

\bibitem[{\citenamefont{Yildiz et~al.}(2008)\citenamefont{Yildiz, Tomishige,
  Gennerich, and Vale}}]{Yildiz2008}
\bibinfo{author}{\bibfnamefont{A.}~\bibnamefont{Yildiz}},
  \bibinfo{author}{\bibfnamefont{M.}~\bibnamefont{Tomishige}},
  \bibinfo{author}{\bibfnamefont{A.}~\bibnamefont{Gennerich}},
  \bibnamefont{and} \bibinfo{author}{\bibfnamefont{R.~D.} \bibnamefont{Vale}},
  \bibinfo{journal}{Cell} \textbf{\bibinfo{volume}{134}}, \bibinfo{pages}{1030}
  (\bibinfo{year}{2008}).

\bibitem[{\citenamefont{Hackney}(2005)}]{Hackney2005}
\bibinfo{author}{\bibfnamefont{D.~D.} \bibnamefont{Hackney}},
  \bibinfo{journal}{Proc. Natl. Acad. Sci. USA} \textbf{\bibinfo{volume}{102}},
  \bibinfo{pages}{18338} (\bibinfo{year}{2005}).

\bibitem[{\citenamefont{Taniguchi et~al.}(2005)\citenamefont{Taniguchi,
  Nishiyama, Ishhi, and Yanagida}}]{Taniguchi2005}
\bibinfo{author}{\bibfnamefont{Y.}~\bibnamefont{Taniguchi}},
  \bibinfo{author}{\bibfnamefont{M.}~\bibnamefont{Nishiyama}},
  \bibinfo{author}{\bibfnamefont{Y.}~\bibnamefont{Ishhi}}, \bibnamefont{and}
  \bibinfo{author}{\bibfnamefont{T.}~\bibnamefont{Yanagida}},
  \bibinfo{journal}{Nat. Chem. Biol.} \textbf{\bibinfo{volume}{1}},
  \bibinfo{pages}{342} (\bibinfo{year}{2005}).

\bibitem[{\citenamefont{Nishiyama et~al.}(2002)\citenamefont{Nishiyama,
  Higuchi, and Yanagida}}]{Nishiyama2002}
\bibinfo{author}{\bibfnamefont{M.}~\bibnamefont{Nishiyama}},
  \bibinfo{author}{\bibfnamefont{H.}~\bibnamefont{Higuchi}}, \bibnamefont{and}
  \bibinfo{author}{\bibfnamefont{T.}~\bibnamefont{Yanagida}},
  \bibinfo{journal}{Nature Cell Biol.} \textbf{\bibinfo{volume}{4}},
  \bibinfo{pages}{790} (\bibinfo{year}{2002}).

\bibitem[{\citenamefont{Schnitzer et~al.}(2000)\citenamefont{Schnitzer,
  Visscher, and Block}}]{Schnitzer2000}
\bibinfo{author}{\bibfnamefont{M.~J.} \bibnamefont{Schnitzer}},
  \bibinfo{author}{\bibfnamefont{K.}~\bibnamefont{Visscher}}, \bibnamefont{and}
  \bibinfo{author}{\bibfnamefont{S.~M.} \bibnamefont{Block}},
  \bibinfo{journal}{Nat. Cell. Biol.} \textbf{\bibinfo{volume}{2}},
  \bibinfo{pages}{718} (\bibinfo{year}{2000}).

\bibitem[{\citenamefont{Block et~al.}(2003)\citenamefont{Block, Asbury,
  Shaevitz, and Lang}}]{Block2003}
\bibinfo{author}{\bibfnamefont{S.~M.} \bibnamefont{Block}},
  \bibinfo{author}{\bibfnamefont{C.~L.} \bibnamefont{Asbury}},
  \bibinfo{author}{\bibfnamefont{J.~W.} \bibnamefont{Shaevitz}},
  \bibnamefont{and} \bibinfo{author}{\bibfnamefont{M.~J.} \bibnamefont{Lang}},
  \bibinfo{journal}{Proc. Natl. Acad. Sci. USA} \textbf{\bibinfo{volume}{100}},
  \bibinfo{pages}{2351} (\bibinfo{year}{2003}).

\bibitem[{\citenamefont{Gennerich et~al.}(2007)\citenamefont{Gennerich, Carter,
  Reck-Peterson, and Vale}}]{Gennerich2007}
\bibinfo{author}{\bibfnamefont{A.}~\bibnamefont{Gennerich}},
  \bibinfo{author}{\bibfnamefont{A.~P.} \bibnamefont{Carter}},
  \bibinfo{author}{\bibfnamefont{S.~L.} \bibnamefont{Reck-Peterson}},
  \bibnamefont{and} \bibinfo{author}{\bibfnamefont{R.~D.} \bibnamefont{Vale}},
  \bibinfo{journal}{Cell} \textbf{\bibinfo{volume}{131}}, \bibinfo{pages}{952}
  (\bibinfo{year}{2007}).

\bibitem[{\citenamefont{Watanabe and Higuchi}(2007)}]{Watanabe2007}
\bibinfo{author}{\bibfnamefont{T.~M.} \bibnamefont{Watanabe}} \bibnamefont{and}
  \bibinfo{author}{\bibfnamefont{H.}~\bibnamefont{Higuchi}},
  \bibinfo{journal}{Biophys. J.} \textbf{\bibinfo{volume}{92}},
  \bibinfo{pages}{4109} (\bibinfo{year}{2007}).

\bibitem[{\citenamefont{Mallik et~al.}(2004)\citenamefont{Mallik, Carter, Lex,
  King, and Gross}}]{Mallik2004}
\bibinfo{author}{\bibfnamefont{R.}~\bibnamefont{Mallik}},
  \bibinfo{author}{\bibfnamefont{B.~C.} \bibnamefont{Carter}},
  \bibinfo{author}{\bibfnamefont{S.~A.} \bibnamefont{Lex}},
  \bibinfo{author}{\bibfnamefont{S.~J.} \bibnamefont{King}}, \bibnamefont{and}
  \bibinfo{author}{\bibfnamefont{S.~P.} \bibnamefont{Gross}},
  \bibinfo{journal}{Nature} \textbf{\bibinfo{volume}{427}},
  \bibinfo{pages}{649} (\bibinfo{year}{2004}).

\bibitem[{\citenamefont{Hirakawa et~al.}(2000)\citenamefont{Hirakawa, Higuchi,
  and Toyoshima}}]{Hirakawa2000}
\bibinfo{author}{\bibfnamefont{E.}~\bibnamefont{Hirakawa}},
  \bibinfo{author}{\bibfnamefont{H.}~\bibnamefont{Higuchi}}, \bibnamefont{and}
  \bibinfo{author}{\bibfnamefont{Y.~Y.} \bibnamefont{Toyoshima}},
  \bibinfo{journal}{Proc. Natl. Acad. Sci. USA} \textbf{\bibinfo{volume}{97}},
  \bibinfo{pages}{2533} (\bibinfo{year}{2000}).

\bibitem[{\citenamefont{Cho et~al.}(2008)\citenamefont{Cho, Reck-Peterson, and
  Vale}}]{Cho2008}
\bibinfo{author}{\bibfnamefont{C.}~\bibnamefont{Cho}},
  \bibinfo{author}{\bibfnamefont{S.~L.} \bibnamefont{Reck-Peterson}},
  \bibnamefont{and} \bibinfo{author}{\bibfnamefont{R.~D.} \bibnamefont{Vale}},
  \bibinfo{journal}{J. Biol. Chem.} \textbf{\bibinfo{volume}{283}},
  \bibinfo{pages}{25839} (\bibinfo{year}{2008}).

\bibitem[{\citenamefont{Ross et~al.}(2008)\citenamefont{Ross, Shuman, Holzbaur,
  and Goldman}}]{Ross2008}
\bibinfo{author}{\bibfnamefont{J.~L.} \bibnamefont{Ross}},
  \bibinfo{author}{\bibfnamefont{H.}~\bibnamefont{Shuman}},
  \bibinfo{author}{\bibfnamefont{E.~L.~F.} \bibnamefont{Holzbaur}},
  \bibnamefont{and} \bibinfo{author}{\bibfnamefont{Y.~E.}
  \bibnamefont{Goldman}}, \bibinfo{journal}{Biophys. J.}
  \textbf{\bibinfo{volume}{94}}, \bibinfo{pages}{3115} (\bibinfo{year}{2008}).

\bibitem[{\citenamefont{King and Schroer}(2000)}]{King2000}
\bibinfo{author}{\bibfnamefont{S.~J.} \bibnamefont{King}} \bibnamefont{and}
  \bibinfo{author}{\bibfnamefont{T.~A.} \bibnamefont{Schroer}},
  \bibinfo{journal}{Nat. Cell. Biol.} \textbf{\bibinfo{volume}{2}},
  \bibinfo{pages}{20} (\bibinfo{year}{2000}).

\bibitem[{\citenamefont{Cappello et~al.}(2007)\citenamefont{Cappello, Pierobon,
  Symonds, Busoni, Gebhardt, Rief, and Prost}}]{Cappello2007}
\bibinfo{author}{\bibfnamefont{G.}~\bibnamefont{Cappello}},
  \bibinfo{author}{\bibfnamefont{P.}~\bibnamefont{Pierobon}},
  \bibinfo{author}{\bibfnamefont{C.}~\bibnamefont{Symonds}},
  \bibinfo{author}{\bibfnamefont{L.}~\bibnamefont{Busoni}},
  \bibinfo{author}{\bibfnamefont{J.~C.~M.} \bibnamefont{Gebhardt}},
  \bibinfo{author}{\bibfnamefont{M.}~\bibnamefont{Rief}}, \bibnamefont{and}
  \bibinfo{author}{\bibfnamefont{J.}~\bibnamefont{Prost}},
  \bibinfo{journal}{Proc. Natl. Acad. Sci. USA} \textbf{\bibinfo{volume}{104}},
  \bibinfo{pages}{15328} (\bibinfo{year}{2007}).

\bibitem[{\citenamefont{Tsygankov and Fisher}(2007)}]{Tsygankov2007}
\bibinfo{author}{\bibfnamefont{D.}~\bibnamefont{Tsygankov}} \bibnamefont{and}
  \bibinfo{author}{\bibfnamefont{M.~E.} \bibnamefont{Fisher}},
  \bibinfo{journal}{Proc. Natl. Acad. Sci. USA} \textbf{\bibinfo{volume}{104}},
  \bibinfo{pages}{19321} (\bibinfo{year}{2007}).

\bibitem[{\citenamefont{Gebhardt et~al.}(2006)\citenamefont{Gebhardt, Clemen,
  Jaud, and Rief}}]{Christof2006}
\bibinfo{author}{\bibfnamefont{J.~C.~M.} \bibnamefont{Gebhardt}},
  \bibinfo{author}{\bibfnamefont{A.~E.-M.} \bibnamefont{Clemen}},
  \bibinfo{author}{\bibfnamefont{J.}~\bibnamefont{Jaud}}, \bibnamefont{and}
  \bibinfo{author}{\bibfnamefont{M.}~\bibnamefont{Rief}},
  \bibinfo{journal}{Proc. Natl. Acad. Sci. USA} \textbf{\bibinfo{volume}{103}},
  \bibinfo{pages}{8680} (\bibinfo{year}{2006}).

\bibitem[{\citenamefont{Clemen et~al.}(2005)\citenamefont{Clemen, Vilfan, Jaud,
  Zhang, B\"{a}rmann, and Rief}}]{Clemen2005}
\bibinfo{author}{\bibfnamefont{A.~E.-M.} \bibnamefont{Clemen}},
  \bibinfo{author}{\bibfnamefont{M.}~\bibnamefont{Vilfan}},
  \bibinfo{author}{\bibfnamefont{J.}~\bibnamefont{Jaud}},
  \bibinfo{author}{\bibfnamefont{J.}~\bibnamefont{Zhang}},
  \bibinfo{author}{\bibfnamefont{M.}~\bibnamefont{B\"{a}rmann}},
  \bibnamefont{and} \bibinfo{author}{\bibfnamefont{M.}~\bibnamefont{Rief}},
  \bibinfo{journal}{Biophys. J.} \textbf{\bibinfo{volume}{88}},
  \bibinfo{pages}{4402} (\bibinfo{year}{2005}).

\bibitem[{\citenamefont{Kolomeisky and Fisher}(2003)}]{Kolomeisky2003}
\bibinfo{author}{\bibfnamefont{A.~B.} \bibnamefont{Kolomeisky}}
  \bibnamefont{and} \bibinfo{author}{\bibfnamefont{M.~E.}
  \bibnamefont{Fisher}}, \bibinfo{journal}{Biophys. J.}
  \textbf{\bibinfo{volume}{84}}, \bibinfo{pages}{1642} (\bibinfo{year}{2003}).

\bibitem[{\citenamefont{Uemura et~al.}(2004)\citenamefont{Uemura, Higuchi,
  Olivares, Cruz, and Ishiwata}}]{Uemura2004}
\bibinfo{author}{\bibfnamefont{S.}~\bibnamefont{Uemura}},
  \bibinfo{author}{\bibfnamefont{H.}~\bibnamefont{Higuchi}},
  \bibinfo{author}{\bibfnamefont{A.~O.} \bibnamefont{Olivares}},
  \bibinfo{author}{\bibfnamefont{E.~M. D.~L.} \bibnamefont{Cruz}},
  \bibnamefont{and} \bibinfo{author}{\bibfnamefont{S.}~\bibnamefont{Ishiwata}},
  \bibinfo{journal}{Nat. Struct. Mol. Biol.} \textbf{\bibinfo{volume}{11}},
  \bibinfo{pages}{877} (\bibinfo{year}{2004}).

\bibitem[{\citenamefont{Oster and Wang}(2000)}]{Oster2000}
\bibinfo{author}{\bibfnamefont{G.}~\bibnamefont{Oster}} \bibnamefont{and}
  \bibinfo{author}{\bibfnamefont{H.}~\bibnamefont{Wang}},
  \bibinfo{journal}{Nature} \textbf{\bibinfo{volume}{32}}, \bibinfo{pages}{459}
  (\bibinfo{year}{2000}).

\bibitem[{\citenamefont{Yardimci et~al.}(2008)\citenamefont{Yardimci, van
  Duffelen, Mao, Rosenfeld, and Selvin}}]{Yardimci2008}
\bibinfo{author}{\bibfnamefont{H.}~\bibnamefont{Yardimci}},
  \bibinfo{author}{\bibfnamefont{M.}~\bibnamefont{van Duffelen}},
  \bibinfo{author}{\bibfnamefont{Y.}~\bibnamefont{Mao}},
  \bibinfo{author}{\bibfnamefont{S.~S.} \bibnamefont{Rosenfeld}},
  \bibnamefont{and} \bibinfo{author}{\bibfnamefont{P.~R.}
  \bibnamefont{Selvin}}, \bibinfo{journal}{Proc. Natl. Acad. Sci. USA}
  \textbf{\bibinfo{volume}{105}}, \bibinfo{pages}{6016} (\bibinfo{year}{2008}).

\bibitem[{\citenamefont{Sweeney and Houdusse}(2007)}]{Sweeney2007}
\bibinfo{author}{\bibfnamefont{H.~L.} \bibnamefont{Sweeney}} \bibnamefont{and}
  \bibinfo{author}{\bibfnamefont{A.}~\bibnamefont{Houdusse}},
  \bibinfo{journal}{Curr. Opin. Cell. Biol.} \textbf{\bibinfo{volume}{19}},
  \bibinfo{pages}{57} (\bibinfo{year}{2007}).

\bibitem[{\citenamefont{Oguchi et~al.}(2008)\citenamefont{Oguchi, Mikhailenko,
  Ohki, Olivares, Cruz, and Ishiwata}}]{Oguchi2008}
\bibinfo{author}{\bibfnamefont{Y.}~\bibnamefont{Oguchi}},
  \bibinfo{author}{\bibfnamefont{S.~V.} \bibnamefont{Mikhailenko}},
  \bibinfo{author}{\bibfnamefont{T.}~\bibnamefont{Ohki}},
  \bibinfo{author}{\bibfnamefont{A.~O.} \bibnamefont{Olivares}},
  \bibinfo{author}{\bibfnamefont{E.~M. D.~L.} \bibnamefont{Cruz}},
  \bibnamefont{and} \bibinfo{author}{\bibfnamefont{S.}~\bibnamefont{Ishiwata}},
  \bibinfo{journal}{Proc. Natl. Acad. Sci. USA} \textbf{\bibinfo{volume}{105}},
  \bibinfo{pages}{7714} (\bibinfo{year}{2008}).

\bibitem[{\citenamefont{Bryant et~al.}(2007)\citenamefont{Bryant, Altman, and
  Spudich}}]{Bryant2007}
\bibinfo{author}{\bibfnamefont{Z.}~\bibnamefont{Bryant}},
  \bibinfo{author}{\bibfnamefont{D.}~\bibnamefont{Altman}}, \bibnamefont{and}
  \bibinfo{author}{\bibfnamefont{J.~A.} \bibnamefont{Spudich}},
  \bibinfo{journal}{Proc. Natl. Acad. Sci. USA} \textbf{\bibinfo{volume}{104}},
  \bibinfo{pages}{772} (\bibinfo{year}{2007}).

\bibitem[{\citenamefont{Iwaki et~al.}(2009)\citenamefont{Iwaki, Iwane,
  Shimokawa, Cooke, and Yanagida}}]{Iwaki2009}
\bibinfo{author}{\bibfnamefont{M.}~\bibnamefont{Iwaki}},
  \bibinfo{author}{\bibfnamefont{A.~H.} \bibnamefont{Iwane}},
  \bibinfo{author}{\bibfnamefont{T.}~\bibnamefont{Shimokawa}},
  \bibinfo{author}{\bibfnamefont{R.}~\bibnamefont{Cooke}}, \bibnamefont{and}
  \bibinfo{author}{\bibfnamefont{T.}~\bibnamefont{Yanagida}},
  \bibinfo{journal}{Nat. Chem. Biol.} \textbf{\bibinfo{volume}{5}},
  \bibinfo{pages}{403} (\bibinfo{year}{2009}).

\bibitem[{\citenamefont{Udovichenko et~al.}(2002)\citenamefont{Udovichenko,
  Gibbs, and Williams}}]{Udovichenko2002}
\bibinfo{author}{\bibfnamefont{I.~P.} \bibnamefont{Udovichenko}},
  \bibinfo{author}{\bibfnamefont{D.}~\bibnamefont{Gibbs}}, \bibnamefont{and}
  \bibinfo{author}{\bibfnamefont{D.~S.} \bibnamefont{Williams}},
  \bibinfo{journal}{J. Cell Sci.} \textbf{\bibinfo{volume}{115}},
  \bibinfo{pages}{445} (\bibinfo{year}{2002}).

\bibitem[{\citenamefont{Inoue et~al.}(2002)\citenamefont{Inoue, Saito, Ikebe,
  and Ikebe}}]{Inoue2002}
\bibinfo{author}{\bibfnamefont{A.}~\bibnamefont{Inoue}},
  \bibinfo{author}{\bibfnamefont{J.}~\bibnamefont{Saito}},
  \bibinfo{author}{\bibfnamefont{R.}~\bibnamefont{Ikebe}}, \bibnamefont{and}
  \bibinfo{author}{\bibfnamefont{M.}~\bibnamefont{Ikebe}},
  \bibinfo{journal}{Nat. Cell. Biol.} \textbf{\bibinfo{volume}{4}},
  \bibinfo{pages}{302} (\bibinfo{year}{2002}).

\bibitem[{\citenamefont{Tominaga et~al.}(2003)\citenamefont{Tominaga, Kojima,
  Yokota, Orii, Nakamori, Katayama, Anson, Shimmen, and Oiwa}}]{Tominaga2003}
\bibinfo{author}{\bibfnamefont{M.}~\bibnamefont{Tominaga}},
  \bibinfo{author}{\bibfnamefont{H.}~\bibnamefont{Kojima}},
  \bibinfo{author}{\bibfnamefont{E.}~\bibnamefont{Yokota}},
  \bibinfo{author}{\bibfnamefont{H.}~\bibnamefont{Orii}},
  \bibinfo{author}{\bibfnamefont{R.}~\bibnamefont{Nakamori}},
  \bibinfo{author}{\bibfnamefont{E.}~\bibnamefont{Katayama}},
  \bibinfo{author}{\bibfnamefont{M.}~\bibnamefont{Anson}},
  \bibinfo{author}{\bibfnamefont{T.}~\bibnamefont{Shimmen}}, \bibnamefont{and}
  \bibinfo{author}{\bibfnamefont{K.}~\bibnamefont{Oiwa}},
  \bibinfo{journal}{EMBO J.} \textbf{\bibinfo{volume}{22}},
  \bibinfo{pages}{1263} (\bibinfo{year}{2003}).

\bibitem[{\citenamefont{Kim et~al.}(2002)\citenamefont{Kim, Narayan, and
  Patel}}]{Kim2002}
\bibinfo{author}{\bibfnamefont{D.-E.} \bibnamefont{Kim}},
  \bibinfo{author}{\bibfnamefont{M.}~\bibnamefont{Narayan}}, \bibnamefont{and}
  \bibinfo{author}{\bibfnamefont{S.~S.} \bibnamefont{Patel}},
  \bibinfo{journal}{J. Mol. Biol.} \textbf{\bibinfo{volume}{321}},
  \bibinfo{pages}{807} (\bibinfo{year}{2002}).

\bibitem[{\citenamefont{Risken}(1989)}]{Risken1989}
\bibinfo{author}{\bibfnamefont{H.}~\bibnamefont{Risken}},
  \emph{\bibinfo{title}{The Fokker-Planck Equation}}
  (\bibinfo{publisher}{Springer, Berlin}, \bibinfo{year}{1989}).

\bibitem[{\citenamefont{Zhang}(2009{\natexlab{b}})}]{Zhang20091}
\bibinfo{author}{\bibfnamefont{Y.}~\bibnamefont{Zhang}}, \bibinfo{journal}{J.
  Stat. Phys.} \textbf{\bibinfo{volume}{134}}, \bibinfo{pages}{669}
  (\bibinfo{year}{2009}{\natexlab{b}}).

\bibitem[{\citenamefont{Wang and Oster}(2002)}]{Wang2002}
\bibinfo{author}{\bibfnamefont{H.}~\bibnamefont{Wang}} \bibnamefont{and}
  \bibinfo{author}{\bibfnamefont{G.}~\bibnamefont{Oster}},
  \bibinfo{journal}{Europhys. Lett.} \textbf{\bibinfo{volume}{57}},
  \bibinfo{pages}{134} (\bibinfo{year}{2002}).

\bibitem[{\citenamefont{Gehlen et~al.}(2008)\citenamefont{Gehlen, Evstigneev,
  and Reimann}}]{Gehlen2008}
\bibinfo{author}{\bibfnamefont{S.~V.} \bibnamefont{Gehlen}},
  \bibinfo{author}{\bibfnamefont{M.}~\bibnamefont{Evstigneev}},
  \bibnamefont{and} \bibinfo{author}{\bibfnamefont{P.}~\bibnamefont{Reimann}},
  \bibinfo{journal}{Phys. Rev. E} \textbf{\bibinfo{volume}{77}},
  \bibinfo{pages}{031136} (\bibinfo{year}{2008}).

\bibitem[{\citenamefont{Nieuwenhuizen et~al.}(2004)\citenamefont{Nieuwenhuizen,
  Klumpp, and Lipowsky}}]{Nieuwenhuizen2004}
\bibinfo{author}{\bibfnamefont{T.~M.} \bibnamefont{Nieuwenhuizen}},
  \bibinfo{author}{\bibfnamefont{S.}~\bibnamefont{Klumpp}}, \bibnamefont{and}
  \bibinfo{author}{\bibfnamefont{R.}~\bibnamefont{Lipowsky}},
  \bibinfo{journal}{Phys. A} \textbf{\bibinfo{volume}{350}},
  \bibinfo{pages}{122} (\bibinfo{year}{2004}).

\bibitem[{\citenamefont{Kolomeisky and Fisher}(2007)}]{Kolomeisky2007}
\bibinfo{author}{\bibfnamefont{A.~B.} \bibnamefont{Kolomeisky}}
  \bibnamefont{and} \bibinfo{author}{\bibfnamefont{M.~E.}
  \bibnamefont{Fisher}}, \bibinfo{journal}{Ann. Rev. Phys. Chem.}
  \textbf{\bibinfo{volume}{58}}, \bibinfo{pages}{675} (\bibinfo{year}{2007}).

\bibitem[{\citenamefont{Liepelt and Lipowsky}(2007)}]{Liepelt2007}
\bibinfo{author}{\bibfnamefont{S.}~\bibnamefont{Liepelt}} \bibnamefont{and}
  \bibinfo{author}{\bibfnamefont{R.}~\bibnamefont{Lipowsky}},
  \bibinfo{journal}{Phys. Rev. Lett.} \textbf{\bibinfo{volume}{98}},
  \bibinfo{pages}{258102} (\bibinfo{year}{2007}).

\bibitem[{\citenamefont{Zhang}(2009{\natexlab{c}})}]{Zhang20093}
\bibinfo{author}{\bibfnamefont{Y.}~\bibnamefont{Zhang}},
  \bibinfo{journal}{Physica A} \textbf{\bibinfo{volume}{383}},
  \bibinfo{pages}{3465} (\bibinfo{year}{2009}{\natexlab{c}}).

\bibitem[{\citenamefont{Fisher and Kolomeisky}(1999)}]{Fisher1999}
\bibinfo{author}{\bibfnamefont{M.~E.} \bibnamefont{Fisher}} \bibnamefont{and}
  \bibinfo{author}{\bibfnamefont{A.~B.} \bibnamefont{Kolomeisky}},
  \bibinfo{journal}{Physica A} \textbf{\bibinfo{volume}{274}},
  \bibinfo{pages}{241} (\bibinfo{year}{1999}).

\bibitem[{\citenamefont{Reimann et~al.}(2001)\citenamefont{Reimann, den Broeck,
  Linke, Hanggi, Rubi, and P\'{e}rez-Madrid}}]{Reimann2001}
\bibinfo{author}{\bibfnamefont{P.}~\bibnamefont{Reimann}},
  \bibinfo{author}{\bibfnamefont{C.~V.} \bibnamefont{den Broeck}},
  \bibinfo{author}{\bibfnamefont{H.}~\bibnamefont{Linke}},
  \bibinfo{author}{\bibfnamefont{P.}~\bibnamefont{Hanggi}},
  \bibinfo{author}{\bibfnamefont{J.~M.} \bibnamefont{Rubi}}, \bibnamefont{and}
  \bibinfo{author}{\bibfnamefont{A.}~\bibnamefont{P\'{e}rez-Madrid}},
  \bibinfo{journal}{Phys. Rev. Lett.} \textbf{\bibinfo{volume}{87}},
  \bibinfo{pages}{010602} (\bibinfo{year}{2001}).

\bibitem[{\citenamefont{Zhang}(2009{\natexlab{d}})}]{Zhang20092}
\bibinfo{author}{\bibfnamefont{Y.}~\bibnamefont{Zhang}},
  \bibinfo{journal}{Phys. Lett. A} \textbf{\bibinfo{volume}{373}},
  \bibinfo{pages}{2629} (\bibinfo{year}{2009}{\natexlab{d}}).

\bibitem[{\citenamefont{Pury and C\'{a}ceres}(2003)}]{Pury2003}
\bibinfo{author}{\bibfnamefont{P.~A.} \bibnamefont{Pury}} \bibnamefont{and}
  \bibinfo{author}{\bibfnamefont{M.~O.} \bibnamefont{C\'{a}ceres}},
  \bibinfo{journal}{J. Phys. A: Math. Gen.} \textbf{\bibinfo{volume}{36}},
  \bibinfo{pages}{2695} (\bibinfo{year}{2003}).

\bibitem[{\citenamefont{Kolomeisky et~al.}(2005)\citenamefont{Kolomeisky,
  Stukalin, and Popov}}]{Kolomeisky2005}
\bibinfo{author}{\bibfnamefont{A.~B.} \bibnamefont{Kolomeisky}},
  \bibinfo{author}{\bibfnamefont{E.~B.} \bibnamefont{Stukalin}},
  \bibnamefont{and} \bibinfo{author}{\bibfnamefont{A.~A.} \bibnamefont{Popov}},
  \bibinfo{journal}{Phys. Rev. E} \textbf{\bibinfo{volume}{71}},
  \bibinfo{pages}{031902} (\bibinfo{year}{2005}).

\bibitem[{\citenamefont{Bieling et~al.}(2008)\citenamefont{Bieling, Telley,
  Piehler, and Surrey}}]{Bieling2008}
\bibinfo{author}{\bibfnamefont{P.}~\bibnamefont{Bieling}},
  \bibinfo{author}{\bibfnamefont{I.~A.} \bibnamefont{Telley}},
  \bibinfo{author}{\bibfnamefont{J.}~\bibnamefont{Piehler}}, \bibnamefont{and}
  \bibinfo{author}{\bibfnamefont{T.}~\bibnamefont{Surrey}},
  \bibinfo{journal}{EMBO Reports} \textbf{\bibinfo{volume}{19}},
  \bibinfo{pages}{1121} (\bibinfo{year}{2008}).

\bibitem[{\citenamefont{Endres et~al.}(2006)\citenamefont{Endres, Yoshioka,
  Milligan, and Vale}}]{Endres2006}
\bibinfo{author}{\bibfnamefont{N.~F.} \bibnamefont{Endres}},
  \bibinfo{author}{\bibfnamefont{C.}~\bibnamefont{Yoshioka}},
  \bibinfo{author}{\bibfnamefont{R.~A.} \bibnamefont{Milligan}},
  \bibnamefont{and} \bibinfo{author}{\bibfnamefont{R.~D.} \bibnamefont{Vale}},
  \bibinfo{journal}{Nature} \textbf{\bibinfo{volume}{439}},
  \bibinfo{pages}{875} (\bibinfo{year}{2006}).

\bibitem[{\citenamefont{Seidel et~al.}(2008)\citenamefont{Seidel, Bloom,
  Dekker, and Szczelkun}}]{Seidel2008}
\bibinfo{author}{\bibfnamefont{R.}~\bibnamefont{Seidel}},
  \bibinfo{author}{\bibfnamefont{J.~G.~P.} \bibnamefont{Bloom}},
  \bibinfo{author}{\bibfnamefont{C.}~\bibnamefont{Dekker}}, \bibnamefont{and}
  \bibinfo{author}{\bibfnamefont{M.~D.} \bibnamefont{Szczelkun}},
  \bibinfo{journal}{EMBO J.} \textbf{\bibinfo{volume}{27}},
  \bibinfo{pages}{1388} (\bibinfo{year}{2008}).

\bibitem[{\citenamefont{Shaevitz et~al.}(2005)\citenamefont{Shaevitz, Block,
  and Schnitzer}}]{Shaevitz2005}
\bibinfo{author}{\bibfnamefont{J.~W.} \bibnamefont{Shaevitz}},
  \bibinfo{author}{\bibfnamefont{S.~M.} \bibnamefont{Block}}, \bibnamefont{and}
  \bibinfo{author}{\bibfnamefont{M.~J.} \bibnamefont{Schnitzer}},
  \bibinfo{journal}{Biophys. J.} \textbf{\bibinfo{volume}{89}},
  \bibinfo{pages}{2277} (\bibinfo{year}{2005}).

\bibitem[{\citenamefont{Gao}(2006)}]{Gao2006}
\bibinfo{author}{\bibfnamefont{Y.~Q.} \bibnamefont{Gao}},
  \bibinfo{journal}{Biophys. J.} \textbf{\bibinfo{volume}{90}},
  \bibinfo{pages}{811} (\bibinfo{year}{2006}).

\bibitem[{\citenamefont{Nishikawa et~al.}(2008)\citenamefont{Nishikawa, Takagi,
  Shibata, Iwane, and Yanagida}}]{Nishikawa2008}
\bibinfo{author}{\bibfnamefont{M.}~\bibnamefont{Nishikawa}},
  \bibinfo{author}{\bibfnamefont{H.}~\bibnamefont{Takagi}},
  \bibinfo{author}{\bibfnamefont{T.}~\bibnamefont{Shibata}},
  \bibinfo{author}{\bibfnamefont{A.~H.} \bibnamefont{Iwane}}, \bibnamefont{and}
  \bibinfo{author}{\bibfnamefont{T.}~\bibnamefont{Yanagida}},
  \bibinfo{journal}{Phys. Rev. Lett.} \textbf{\bibinfo{volume}{101}},
  \bibinfo{pages}{128103} (\bibinfo{year}{2008}).

\bibitem[{\citenamefont{Masuda}(2009)}]{Masuda2009}
\bibinfo{author}{\bibfnamefont{T.}~\bibnamefont{Masuda}},
  \bibinfo{journal}{BioSystems} \textbf{\bibinfo{volume}{95}},
  \bibinfo{pages}{104} (\bibinfo{year}{2009}).

\bibitem[{\citenamefont{Gerritsma and Gaspard}(2009)}]{Gerritsma2009}
\bibinfo{author}{\bibfnamefont{E.}~\bibnamefont{Gerritsma}} \bibnamefont{and}
  \bibinfo{author}{\bibfnamefont{P.}~\bibnamefont{Gaspard}},
  \bibinfo{journal}{arXiv:0904.4218}  (\bibinfo{year}{2009}).

\bibitem[{\citenamefont{Lipowsky}(2000)}]{Lipowsky2000}
\bibinfo{author}{\bibfnamefont{R.}~\bibnamefont{Lipowsky}},
  \bibinfo{journal}{Phys. Rev. Lett.} \textbf{\bibinfo{volume}{85}},
  \bibinfo{pages}{4401} (\bibinfo{year}{2000}).

\bibitem[{\citenamefont{Lipowsky and Jaster}(2003)}]{Lipowsky2003}
\bibinfo{author}{\bibfnamefont{R.}~\bibnamefont{Lipowsky}} \bibnamefont{and}
  \bibinfo{author}{\bibfnamefont{N.}~\bibnamefont{Jaster}},
  \bibinfo{journal}{J. Stat. Phys.} \textbf{\bibinfo{volume}{110}},
  \bibinfo{pages}{1141} (\bibinfo{year}{2003}).

\bibitem[{\citenamefont{Chen et~al.}(1999)\citenamefont{Chen, Yan, and
  Miura}}]{Chen1999}
\bibinfo{author}{\bibfnamefont{Y.}~\bibnamefont{Chen}},
  \bibinfo{author}{\bibfnamefont{B.}~\bibnamefont{Yan}}, \bibnamefont{and}
  \bibinfo{author}{\bibfnamefont{R.}~\bibnamefont{Miura}},
  \bibinfo{journal}{Phys. Rev. E} \textbf{\bibinfo{volume}{60}},
  \bibinfo{pages}{3771} (\bibinfo{year}{1999}).

\bibitem[{\citenamefont{Wang et~al.}(2003)\citenamefont{Wang, Peskin, and
  Elston}}]{Wang2003}
\bibinfo{author}{\bibfnamefont{H.~Y.} \bibnamefont{Wang}},
  \bibinfo{author}{\bibfnamefont{C.~S.} \bibnamefont{Peskin}},
  \bibnamefont{and} \bibinfo{author}{\bibfnamefont{T.~C.}
  \bibnamefont{Elston}}, \bibinfo{journal}{J. theor. Biol.}
  \textbf{\bibinfo{volume}{221}}, \bibinfo{pages}{491} (\bibinfo{year}{2003}).

\bibitem[{\citenamefont{Wang}(2004)}]{Wang2004}
\bibinfo{author}{\bibfnamefont{H.}~\bibnamefont{Wang}}, \bibinfo{journal}{Int.
  J. Numer. Anal. Model.} \textbf{\bibinfo{volume}{1}}, \bibinfo{pages}{1}
  (\bibinfo{year}{2004}).

\bibitem[{\citenamefont{Astumian}(1997)}]{Astumian1997}
\bibinfo{author}{\bibfnamefont{R.~D.} \bibnamefont{Astumian}},
  \bibinfo{journal}{Science} \textbf{\bibinfo{volume}{276}},
  \bibinfo{pages}{917} (\bibinfo{year}{1997}).

\bibitem[{\citenamefont{Parmeggiani et~al.}(1999)\citenamefont{Parmeggiani,
  J\"{u}licher, Ajdari, and Prost}}]{Parmeggiani1999}
\bibinfo{author}{\bibfnamefont{A.}~\bibnamefont{Parmeggiani}},
  \bibinfo{author}{\bibfnamefont{F.}~\bibnamefont{J\"{u}licher}},
  \bibinfo{author}{\bibfnamefont{A.}~\bibnamefont{Ajdari}}, \bibnamefont{and}
  \bibinfo{author}{\bibfnamefont{J.}~\bibnamefont{Prost}},
  \bibinfo{journal}{Physical Review E} \textbf{\bibinfo{volume}{60}},
  \bibinfo{pages}{2127} (\bibinfo{year}{1999}).

\bibitem[{\citenamefont{Parrondo and Cisneros}(2002)}]{Parrondo2002}
\bibinfo{author}{\bibfnamefont{J.~M.~R.} \bibnamefont{Parrondo}}
  \bibnamefont{and} \bibinfo{author}{\bibfnamefont{B.~J.~D.}
  \bibnamefont{Cisneros}}, \bibinfo{journal}{Appl. Phys. A}
  \textbf{\bibinfo{volume}{75}}, \bibinfo{pages}{179} (\bibinfo{year}{2002}).

\bibitem[{\citenamefont{Reimann}(2002)}]{Reimann20021}
\bibinfo{author}{\bibfnamefont{P.}~\bibnamefont{Reimann}},
  \bibinfo{journal}{Phys. Rep.} \textbf{\bibinfo{volume}{361}},
  \bibinfo{pages}{57} (\bibinfo{year}{2002}).

\bibitem[{\citenamefont{Bier and Astumian}(1993)}]{Bier1993}
\bibinfo{author}{\bibfnamefont{M.}~\bibnamefont{Bier}} \bibnamefont{and}
  \bibinfo{author}{\bibfnamefont{R.~D.} \bibnamefont{Astumian}},
  \bibinfo{journal}{Phys. Rev. Lett.} \textbf{\bibinfo{volume}{71}},
  \bibinfo{pages}{1649} (\bibinfo{year}{1993}).

\bibitem[{\citenamefont{J\"{u}licher and Prost}(1995)}]{Frank1995}
\bibinfo{author}{\bibfnamefont{F.}~\bibnamefont{J\"{u}licher}}
  \bibnamefont{and} \bibinfo{author}{\bibfnamefont{J.}~\bibnamefont{Prost}},
  \bibinfo{journal}{Phys. Rev. Lett.} \textbf{\bibinfo{volume}{75}},
  \bibinfo{pages}{2618} (\bibinfo{year}{1995}).

\bibitem[{\citenamefont{Prost et~al.}(1994)\citenamefont{Prost, Chauwin,
  Peliti, and Ajdari}}]{Prost1994}
\bibinfo{author}{\bibfnamefont{J.}~\bibnamefont{Prost}},
  \bibinfo{author}{\bibfnamefont{J.-F.} \bibnamefont{Chauwin}},
  \bibinfo{author}{\bibfnamefont{L.}~\bibnamefont{Peliti}}, \bibnamefont{and}
  \bibinfo{author}{\bibfnamefont{A.}~\bibnamefont{Ajdari}},
  \bibinfo{journal}{Phys. Rev. Lett.} \textbf{\bibinfo{volume}{72}},
  \bibinfo{pages}{2652} (\bibinfo{year}{1994}).

\bibitem[{\citenamefont{Zhang}(2010)}]{Zhang2010}
\bibinfo{author}{\bibfnamefont{Y.}~\bibnamefont{Zhang}},
  \bibinfo{journal}{Chin. J. Chem. Phys.} \textbf{\bibinfo{volume}{23}},
  \bibinfo{pages}{65} (\bibinfo{year}{2010}).

\bibitem[{\citenamefont{Derrida}(1983)}]{Derrida1983}
\bibinfo{author}{\bibfnamefont{B.}~\bibnamefont{Derrida}}, \bibinfo{journal}{J.
  Stat. Phys.} \textbf{\bibinfo{volume}{31}}, \bibinfo{pages}{433}
  (\bibinfo{year}{1983}).

\bibitem[{\citenamefont{Qian}(1997)}]{Qian1997}
\bibinfo{author}{\bibfnamefont{H.}~\bibnamefont{Qian}},
  \bibinfo{journal}{Biophys. Chem.} \textbf{\bibinfo{volume}{67}},
  \bibinfo{pages}{263} (\bibinfo{year}{1997}).

\end{thebibliography}
\end{document}